\documentclass[11pt,a4paper,epsf,epsfig,psfrag]{article}
\usepackage{jheppub}
\usepackage{amsmath,graphicx,amsfonts, mathrsfs,amssymb}
\usepackage{amsmath,epsfig}
\usepackage{amssymb,amsfonts}
\usepackage{latexsym}
\usepackage{subfigure}
\usepackage{pdfsync}
\usepackage{braket}
\usepackage{bm}

\usepackage{latexsym}
\usepackage{color}
\newbox\pippobox
\def\be{\begin{equation}}
\def\ee{\end{equation}}
\def\bea{\begin{eqnarray}}
\def\eea{\end{eqnarray}}

\def\ee           {{\rm e}}

\newcommand{\beq}{\begin{equation}}
\newcommand{\eeq}{\end{equation}}
\newcommand{\beqa}{\begin{eqnarray}}
\newcommand{\eeqa}{\end{eqnarray}}
\newcommand{\beqar}{\begin{eqnarray*}}
\newcommand{\eeqar}{\end{eqnarray*}}

\renewcommand{\eqref}[1]{(\ref{#1})}

\catcode`\@=12

\title{R\'enyi entropy of locally excited states with thermal and boundary effect in 2D CFTs}
\author[a,b]{Wu-Zhong Guo,}
\author[b,a]{Song He,}

\affiliation[a]{State Key Laboratory of Theoretical Physics,
Institute of Theoretical Physics, Chinese Academy of Science,
Beijing 100190, P. R. China }
\affiliation[b]{Yukawa Institute for Theoretical Physics, Kyoto University, Kitashirakawa Oiwakecho,
Sakyo-ku, Kyoto 606-8502, Japan}
\emailAdd{wuzhong@itp.ac.cn}\emailAdd{hesong17@gmail.com}

\abstract{We study R\'enyi entropy of locally excited states with considering the thermal and boundary effects respectively in two dimensional conformal
field theories (CFTs). Firstly, we consider locally excited states obtained by acting primary operators on a thermal state in low temperature limit. The R\'enyi entropy is summation of contribution from thermal effect and local excitation. Secondly, we mainly study the R\'enyi entropy of locally excited states in 2D CFT with a boundary. We show that the time evolution of R\'enyi entropy is affected by the boundary, but does not depend on the boundary condition. Moreover, we show that the maximal value of R\'enyi entropy always coincides with the log of quantum dimension of the primary operator. In terms of quasi-particle interpretation, the boundary behaves as an infinite potential barrier which reflects any energy moving towards it. }


\begin{document}
\maketitle
\section{Introduction}

Many kinds of observables can be defined in Quantum field theories (QFTs).
When we study global or non-local structures, entanglement entropy (EE) or the entanglement R\'enyi entropy (RE) are very helpful quantities. For a subsystem $A$, both of them are defined as a function of the reduced density matrix $\rho_A$. The reduced density matrix $\rho_A$ can be defined from the original density matrix $\rho$ by tracing out the subsystem $B$ which is the complementary of $A$.

One may be curious about whether there is a kind of topological contribution in entanglement entropy even for gapless theories, particularly for conformal field theories (CFTs). For example, topological properties can be qualified by computing topological contributions in entanglement entropy called topological entanglement entropy \cite{wen}. In this paper, we focus on extracting such kind of topological quantity from both R\'enyi entropy and von-Neumann entropy of locally excited states in two dimensional rational CFTs with thermal effect and boundary effect.

The $n$-th R\'enyi entanglement entropy is defined by $S^{(n)}_A=\log\mbox{Tr}[\rho_A^n]/(1-n)$ formally.
The limit $n\to 1$ coincides with the von-Neumann entropy. This is standard replica trick method to calculate the entanglement entropy.
The difference of $S^{(n)}_A$ between the locally excited states and the ground states with introducing thermal and boundary effects are main interest in this paper. The difference is denoted by $\Delta S^{(n)}_A$. Replica approach calculations of $\Delta S^{(n)}_A$ for states excited by operators have been given in \cite{UAM,Nozaki:2014uaa,Nozaki:2014hna}. We will closely follow the construction in \cite{Nozaki:2014hna}\cite{He:2014mwa}\cite{Nozaki:2014uaa}with introducing thermal      and boundary effect.

We would like to review the thermal effect and boundary effect respectively. There are many studies about thermal effect in 2D CFTs. For thermal states, EE is not a good measurable quantity, which is contaminated by the thermal entropy of the subregion. In high temperature limit, the EE will be dominated by thermal entropy. To reveal the quantum entanglement of system with thermal effect, one should identify the thermal contribution and other contributions of EE. Ref. \cite{Herzog:2012bw}conjectured universal form of correction of EE in any quantum system with mass gap. Ref.\cite{Cardy:2014jwa} provided the form of the coefficient of such correction in 2D CFT. To generalize studies in \cite{Cardy:2014jwa} to higher dimensions, Ref.\cite{Herzog:2014fra} and Ref.\cite{Herzog:2014tfa} considered thermal corrections to the entanglement entropy on spheres. On the other hand, the dynamics in 2D CFTs with a boundary have many new features  comparing with 2D CFTs in full complex plane. The original works have been done by Cardy, who discussed surface critical behavior of correlation functions \cite{cardy1}. Ref.\cite{cardy2} studied the constraints on the operator content by imposing by boundary conditions and also the classification of boundary states in terms of the modular transformation. In \cite{cardy3}\cite{cardylewellen}, the concept of boundary operators have been introduced. Ref.\cite{3sp2} showed that the resulting set of boundary conditions to be complete. There are also nice correspondence called as AdS/BCFT proposed by \cite{Takayanagi:2011zk}\cite{Fujita:2011fp}. The boundary effect can be also studied holographically, which is beyond the issue considered in this paper.

In 2D rational CFTs, the authors in \cite{He:2014mwa} get an amazing result for the locally excited states, which relates the R\'enyi entropy to the quantum dimension of the primary operator which is kind of topological quantity. In this paper, we generalize the previous study \cite{He:2014mwa} on R\'enyi entropy with thermal and boundary effects. Firstly, there is a simple sum rule between the thermal correction and local excitation in low temperature limit. That is to say the total R\'enyi entropy are summing over R\'enyi entropy of local excitation and the one of thermal excitation in low temperature limit. Such kind of relation is similar to the sum rule related to the R\'enyi entropy in \cite{Nozaki:2014uaa}. One can generalize the result to local excitation in pure state in 2D CFT. We make use of a different approach \cite{Herzog:2012bw} to obtain the thermal correction to R\'enyi entropy which can be reduced to \cite{Caputa:2014eta}. Secondly, we investigate the the R\'enyi entropy for states excited by local primary operators in the rational CFTs with a boundary. These boundaries introduced here do not break the conformal symmetry. Such theories are called as BCFTs. We show the time evolution of  the R\'enyi entropy in 2D free field theory and Ising model. Then we generalize to rational CFTs with a boundary. The boundary changes the time evolution of the R\'enyi entropy, but does not change the maximal value of the R\'enyi entropy. All these cases studied in this paper show that the R\'enyi entropy does not depend on the choice of boundary conditions. In 2D rational CFTs with a boundary, we also show that the maximal value of the R\'enyi entropy always coincides with the log of quantum dimension of the primary operator during some periods of the evolution. We give the physical understanding of boundary effect, which support the quasi-particles explanation of the local excitation. The boundary behaves as an infinite potential barrier which reflects the quasi-particle moving towards it.

The layout of this paper is as follows. In section 2, we study the thermal effect on the R\'enyi entropy of the local excited state in low temperature limit. In section 3, we set up the local excitation in 2D CFT with a boundary and obtain the R\'enyi entropy of a subsystem with time evolution. We study the 2D free scalar, and Ising model as examples, then generalize the result to 2D rational CFT. In section 4, we devote to the conclusion and physical interpretations of such kinds of effects shown in this paper.
\section{Local excitation in non-vacuum states}

\begin{figure}[h]
\begin{center}
\epsfxsize=8.0 cm \epsfysize=5.0 cm \epsfbox{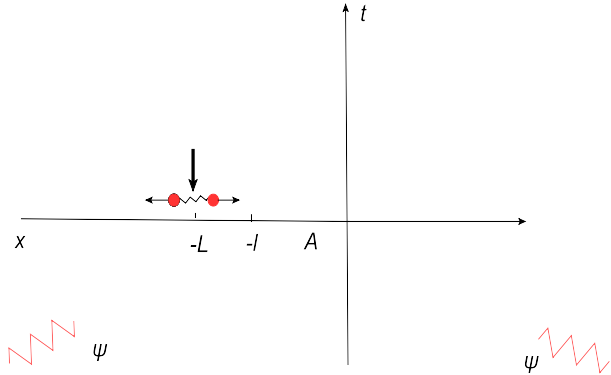}
\end{center}
\caption{This figure is to show our setup in two dimensional complex plane $\omega= x+i t$. The system will be triggered at $x=-L$ and there are left- and right-moving quasi-particle at $t=0$. This setup is same as the one in 2D CFTs without boundary \cite{He:2014mwa}.}\label{fig0}
\end{figure}

In this section, we would like to study the local excitation of thermal state.
We consider a system with temperature $T=1/\beta$ and assume the excitation is local at $x=-L$ by primary operator $O$ shown in fig.[\ref{fig0}]. In this section, we just only consider the low temperature case with large $\beta$ \cite{Herzog:2014fra}. The subsystem $A$ is $-l<x<0$. The density matrix $\rho(t)$ is
\begin{eqnarray}\label{T1}
 \rho(t)=N(t)\Big(O(\omega_2,\bar \omega_2)(\sum \Ket{n}\Bra{n} e^{-\beta E_n-2\beta \epsilon E_n})O^{\dag}(\omega_1,\bar \omega_1)\Big),
\end{eqnarray}
where we have considered the real time evolution, $\epsilon$ is the ultraviolet regularization, $N(t)$ is fixed by normalization condition $tr\rho(t)=1$, $E_n$ are the energy of the excited states. The complex coordinates in $\omega$ plane are listed as follows.
\begin{eqnarray}
 \omega_1=i({\epsilon-it})-L, \ \ \omega_2=-i(\epsilon+it)-L, \nonumber \\
\bar \omega_1=-i(\epsilon-it)-L,\ \  \bar \omega_2=i(\epsilon+it)-L.
\end{eqnarray}
For later convenience, we define
\begin{eqnarray}
&&\rho_0(t)=tr_B( O(\omega_2,\bar \omega_2)\Ket{0}\Bra{0}O^{\dag}(\omega_1,\bar \omega_1)),\nonumber\\
&&\rho_1(t)=tr_B( e^{-\beta E_1-2\beta \epsilon E_1} O(\omega_2,\bar \omega_2)\Ket{1}\Bra{1}O^{\dag}(\omega_1,\bar \omega_1)),
\end{eqnarray}
which can be taken as the reduced density matrix related to the vacuum and first excited state respectively, where we normalize the vacuum energy to be zero, $B$ is the complementary part of subsystem $A$. In the low temperature expansion with $\beta E_1 \ll 1$
\begin{eqnarray}\label{Expand}
\rho_A(t)=tr_B \rho(t)=\frac{\rho_0(t)+\rho_1(t)+...}{tr_A (\rho_0(t)+\rho_1(t))+...}.
\end{eqnarray}
In terms of the definition of the R\'enyi entropy
\begin{eqnarray}\label{resultofthermal}
S^{(n)}_A&=&\frac{\log tr \rho_A(t)^n }{1-n}\nonumber \\ &&\simeq \frac{1}{1-n}\log\Big[\frac{tr(\rho_0(t)^n)}{(tr\rho_0(t))^n}(1+\frac{n tr(\rho_0(t)^{n-1}\rho_1(t))}{tr(\rho_0(t)^n)}-\frac{n tr\rho_1(t)}{tr\rho_0(t)})\Big]\nonumber \\
&&\simeq \frac{1}{1-n}\Big[\log \frac{tr(\rho_0(t)^n)}{(tr\rho_0(t))^n}+\frac{n tr(\rho_0(t)^{n-1}\rho_1(t))}{tr(\rho_0(t)^n)}-\frac{n tr\rho_1(t)}{tr\rho_0(t)}\Big].
\end{eqnarray}
When there is no local excitation, i.e., the operator $O=I$, the result is the same as \cite{Cardy:2014jwa}. The second and third terms of the last line in (\ref{resultofthermal}) involve in the coupling between the local excitation and thermal environment.
In terms of the state operator correspondence, one can denote $\Ket{1}=\lim_{t\to -\infty}\psi(x,t)\Ket{0}$ \cite{Cardy:2014jwa}for the excited state with energy $E_1$. Here we consider the whole system has an infrared cut-off $\Lambda$, and $l/\Lambda \ll 1$. Using the path integral language, (\ref{resultofthermal}) is
\begin{eqnarray}\label{ResultSncorrelation}
S^{(n)}_A&=&\frac{1}{1-n} \log \frac{\langle O^{\dag}(\omega_1,\bar \omega_1)O(\omega_2,\bar \omega_2)...O(\omega_{2n},\bar \omega_{2n})\rangle_{C_n}}{(\langle O^{\dag}(\omega_1,\bar \omega_1)O(\omega_2,\bar \omega_2)\rangle_{C_1})^n}\nonumber \\
&+&\frac{ne^{-\beta E_1}}{1-n}\frac{\langle O^{\dag}(\omega_1,\bar \omega_1)O(\omega_2,\bar \omega_2)...O(\omega_{2n},\bar \omega_{2n})\psi(-\infty)\psi(+\infty)\rangle_{C_n}}{\langle O^{\dag}(\omega_1,\bar \omega_1)O(\omega_2,\bar \omega_2)...O(\omega_{2n},\bar \omega_{2n})\rangle_{C_n}}\nonumber \\
&-&\frac{n e^{-\beta E_1}}{1-n}\frac{\langle O^{\dag}(\omega_1,\bar \omega_1)O(\omega_2,\bar \omega_2) \psi(-\infty)\psi(+\infty)\rangle_{C_1}}{\langle O^{\dag}(\omega_1,\bar \omega_1)O(\omega_2,\bar \omega_2)\rangle_{C_1}},
\end{eqnarray}
where $C_n$ is the $n$-sheet cylinder with circumference $\Lambda$ and $O(\omega_1,\bar \omega_1)...O(\omega_{2n},\bar \omega_{2n})$ are the operators that are inserted in the suitable place in the $n$-sheet cylinder.  The first term (\ref{ResultSncorrelation}) is given by \cite{He:2014mwa} for the local excitation in vacuum. We will study the second and third term of (\ref{ResultSncorrelation}) in detail.

The following transformation \cite{cag} can map the $n$-sheet cylinder to a cylinder with circumference $\Lambda$,
\begin{eqnarray}\label{thermaltransformation}
z=\Big(\frac{e^{2i\pi \omega/\Lambda}-1}{e^{2i\pi \omega/\Lambda}-e^{2i\pi l/\Lambda}}\Big)^{1/n}.
\end{eqnarray}
The points $\omega=-\infty$ and $\omega=\infty$ are mapping to $z_{-\infty}=e^{-2i\pi l/\Lambda}$ and  $z_{+\infty}=1$ respectively. For simplifying our analysis, we only consider the range $|\omega|\ll \Lambda$. Otherwise, the following calculation will be much more complicated.  But the final statement does not change without this approximation.  Thus (\ref{thermaltransformation}) reduces to
\begin{eqnarray}
 z\simeq \Big(\frac{w}{w-l}\Big)^{1/n},
\end{eqnarray}
which is same as the one that is used in \cite{He:2014mwa}.  The points $z_1,...,z_{2n}$ are given by
\begin{eqnarray}
z_{2k+1}&=&e^{2i\pi k/n}\Big(\frac{i\epsilon +t-L}{i\epsilon +t-L-l} \Big)^{1/n},\nonumber \\
 z_{2k+2}&=&e^{2i\pi k/n}\Big(\frac{-i\epsilon +t-L}{-i\epsilon +t-L-l} \Big)^{1/n},\nonumber \\
\bar z_{2k+1}&=&e^{2i\pi k/n}\Big(\frac{-i\epsilon-t-L}{-i\epsilon -t-L-l} \Big)^{1/n},\nonumber \\
\bar z_{2k+2}&=&e^{2i\pi k/n}\Big(\frac{i\epsilon -t-L}{i\epsilon -t-L-l} \Big)^{1/n}.\label{transformationrule}
\end{eqnarray}
In $t>l+L$ or $0<t<L$,
\begin{eqnarray}\label{factor1thermal}
z_{2k+1}-z_{2k+2}\simeq -\frac{2iL\epsilon}{n(t-L)(t-l-L)}z_{2k+1},\nonumber \\
\bar z_{2k+1}-\bar z_{2k+2}\simeq -\frac{2iL\epsilon}{n(t+L)(t-l+L)}\bar z_{2k+1}.
\end{eqnarray}
In $L<t<L+l$,
\begin{eqnarray}\label{factor1therma2}
 z_{2k}-z_{2k+1}\simeq -\frac{2iL\epsilon}{n(t-L)(t-l-L)}z_{2k},\nonumber \\
\bar z_{2k+1}-\bar z_{2k+2}\simeq -\frac{2iL\epsilon}{n(t+L)(t-l+L)}\bar z_{2k+1}.
\end{eqnarray}
In $t>l+L$ or $0<t<L$ the second term of (\ref{ResultSncorrelation}) is
 \begin{eqnarray}
 &&\frac{n e^{-\beta E_1}}{1-n}\frac{\langle O^{\dag}(\omega_1,\bar \omega_1)O(\omega_2,\bar \omega_2)...O(\omega_{2n},\bar \omega_{2n})\psi(-\infty)\psi(+\infty)\rangle_{C_n}}{\langle O^{\dag}(\omega_1,\bar \omega_1)O(\omega_2,\bar \omega_2)...O(\omega_{2n},\bar \omega_{2n})\rangle_{C_n}}\nonumber \\
&=&\frac{ne^{-\beta E_1}}{1-n}\frac{\langle O^{\dag}(z_1,\bar z_1)O(z_2,\bar z_2)...O(z_{2n},\bar z_{2n})\psi^{'}(z_{-\infty})\psi^{'}(z_{+\infty})\rangle_{C_1}}{\langle O^{\dag}(z_1,\bar z_1)O(z_2,\bar z_2)...O(z_{2n},\bar z_{2n})\rangle_{C_1}}\nonumber \\
&=&\frac{ne^{-\beta E_1}}{1-n}\langle \psi^{'}(z_{-\infty})\psi^{'}(z_{+\infty})\rangle_{C_1}.
\end{eqnarray}
In the above formula, we label  $\psi^{'}(z,\bar z)$ as the map of the operator $\psi(\omega,\bar \omega)$\footnote{The operator may not be a primary operator , such as the energy-momentum tensor $T$. A Schwarzian derivative term related to energy momentum tensor operator \cite{Cardy:2014jwa}\cite{Chen:2014unl} will present due to conformal transformation.}. In the second step due to (\ref{factor1thermal}) we have used
 \begin{eqnarray}\label{Thermalcasefactorize}
 &&\langle O^{\dag}(z_1,\bar z_1)O(z_2,\bar z_2)...O(z_{2n},\bar z_{2n})\psi^{'}(z_{-\infty})\psi^{'}(z_{+\infty})\rangle_{C_1}\nonumber \\
 &&\simeq \langle O^{\dag}(z_1,\bar z_1)O(z_2,\bar z_2)\rangle...\langle O^{\dag}(z_{2n-1},\bar z_{2n-1}) O(z_{2n},\bar z_{2n})\rangle \langle \psi^{'}(z_{-\infty})\psi^{'}(z_{+\infty})\rangle.\nonumber\\
 \end{eqnarray}

In $L<t<L+l$, the correlation function can not be factorized as (\ref{Thermalcasefactorize}) directly due to (\ref{factor1therma2}). Following logic in \cite{He:2014mwa}, one can make use of $n-1$ times fusion transformation $(z_1,z_2)(z_3,z_4)...(z_{2n-1},z_{2n}) \to (z_2,z_3)(z_4,z_5)...(z_{2n},z_1)$. The second terms of (\ref{ResultSncorrelation}) is still given by in $\epsilon \rightarrow 0$
\begin{eqnarray}
\frac{ne^{-\beta E_1}}{1-n}\langle \psi^{'}(z_{-\infty})\psi^{'}(z_{+\infty})\rangle_{C_1}.
\end{eqnarray} Here we assume that there are no nontrival correlation between $O$ and $\psi$ for simplifying analysis.
The third term of (\ref{ResultSncorrelation}) in the limit $\epsilon \to 0$ is
\begin{eqnarray}
&&-\frac{ne^{-\beta E_1}}{1-n}\frac{\langle O^{\dag}(\omega_1,\bar \omega_1)O(\omega_2,\bar \omega_2) \psi(-\infty)\psi(+\infty)\rangle_{C_1}}{\langle O^{\dag}(\omega_1,\bar \omega_1)O(\omega_2,\bar \omega_2)\rangle_{C_1}}\nonumber \\
&=&-\frac{n}{1-n}\langle \psi(-\infty)\psi(+\infty)\rangle_{C_1}=-\frac{n}{1-n}e^{-\beta E_1}.
\end{eqnarray}
The sum of the second and third term is the same as the thermal correlation \cite{Cardy:2014jwa} for the short interval limit.
(\ref{ResultSncorrelation}) is the summation over the thermal correction of the vacuum state and local excitation in the vacuum state in low temperature limit. The R\'enyi entropy is summation of thermal effect  and local excitation.
This relationship can be considered as a sum rule for R\'enyi entropy of low temperature thermal state with local excitation, which is different with the sum rule proposed in \cite{Nozaki:2014hna}. For the pure state one could also get a similar sum rule.  The local excitation in thermal states have been also studied in \cite{Caputa:2014eta} for free boson and Ising model.  With a different method \cite{Cardy:2014jwa}, we can reproduce the R\'enyi entropy \cite{Caputa:2014eta} for short interval in the low temperature with taking limit $\epsilon \to 0$ for general 2D rational CFT.

\section{Local excitation in 2D CFTs with a boundary}
In this section, we would like to study the R\'enyi entropy of locally excited state in 2D CFT with a boundary which preserves conformal symmetry. The global property of the CFT with a boundary has been discussed in literature\cite{cag}\cite{Cardy1}\cite{Cardy2}. As we know, the R\'enyi entropy is sensitive to correlation function. The boundary will change the correlation functions. And the boundary conditions also affect the correlation functions.  It is interesting to check what will happen to the R\'enyi entropy for the local excitation in 2D CFT with a boundary.

\subsection{Set-up of local exciation}\label{3.1}
We begin with a CFT with a boundary at $x=0$ and the CFT is living in the range $x\le 0$. We divide the this region into two parts, one part is $-l<x<0$ denoted by A and the other is complement to the region A denoted by B. The time $t$ vary from $-\infty$ to $+\infty$, and the Hamiltonian $H$ is well defined as a operator to generate the time evolution.

We assume that the local excitation of vacuum is at $x=-L$ and consider the R\'enyi entropy of the subsystem A. The time dependent density matrix can be written as
\begin{eqnarray}
\rho(t)=N O(\omega_2,\bar \omega_2)\Ket{0}\Bra{0}O^{\dag}(\omega_1,\bar \omega_1),
\end{eqnarray}
where the coordinates are,
\begin{eqnarray}
\omega_1=i(\epsilon-it)-L, \ \ \omega_2=-i(\epsilon+it)-L, \nonumber \\
\bar \omega_1=-i(\epsilon-it)-L,\ \  \bar \omega_2=i(\epsilon+it)-L.
\end{eqnarray}

\begin{figure}[h]
\begin{center}
\epsfxsize=6.0 cm \epsfysize=5.0 cm \epsfbox{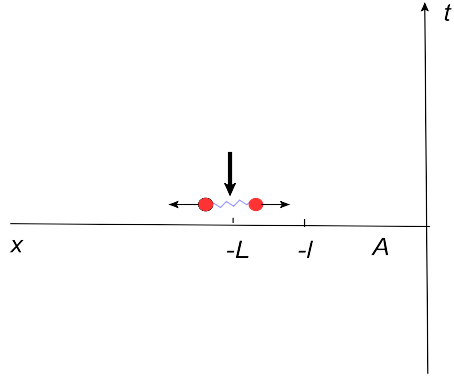}
\end{center}
\caption{This figure is to show our setup in two dimensional left half plane $\omega=x+i t$ with a boundary $x=0$. The system will be triggered at $x=-L$ and there are left- and right-moving quasi-particle at $t=0$.}\label{fig1}
\end{figure}
We still make use of replica trick to study the variation of the R\'enyi entropy of the subsystem A in this section. By definition, the variance of $n$-th R\'enyi entropy can be calculated as
\begin{eqnarray}\label{Mainformula}
\Delta S^{(n)}_A&=&\frac{1}{1-n} \Big[\log \langle O^{\dag}(\omega_1,\bar \omega_1)O(\omega_2,\bar \omega_2)...O(\omega_{2n},\bar \omega_{2n})\rangle_{B_n}\nonumber \\
&-&n\log \langle O^{\dag}(\omega_1,\bar \omega_1)O(\omega_2,\bar \omega_2)\rangle_{B_1}\Big],
\end{eqnarray}
where $B_n$ is the $n$-sheet Riemann surface that consists of n copies of original plane $x\le 0$ with gluing together along $-l \leqslant x \leqslant 0$, $t=0$. With the following conformal transformation,
\begin{eqnarray}\label{Transformation1}
z^n=\frac{\omega+l}{\omega-l}
\end{eqnarray}
the $n$-sheet Riemann surface can be mapped to a disc $|z|\leqslant 1$ which is smooth surface. The boundary x=0 corresponds to $|z|=1$. Furthermore, we can map disc to the upper half plane (UHP) $t \geqslant 0$ with the other conformal map
\begin{eqnarray}
\xi=-i\frac{z+1}{z-1}\label{Transformation2}.
\end{eqnarray}
After two conformal maps, one can make use of well known results of 2D CFT on UHP. Finally, the variation of the R\'enyi entropy (\ref{Mainformula}) is
\begin{eqnarray}\label{result1}
\Delta S_{A}^{(n)}=\frac{1}{1-n}\log \Big[\prod_{k=1}^{2n} (\frac{d\omega_k}{d\xi_k})^{-h}(\frac{d\omega_k}{d\xi_k})^{-\bar h}\frac{\langle O^{\dag}(\xi_1,\bar \xi_1)O(\xi_2,\bar \xi_2)...O^{\dag}(\xi_{2n-1},\bar \xi_{2n-1})O(\xi_{2n},\bar \xi_{2n})\rangle_{UHP}}{(\langle O^{\dag}(\omega_1,\bar \omega_1)O(\omega_2,\bar \omega_2)\rangle_{B_1})^n}\Big].\nonumber\\
\
\end{eqnarray}
This is the key formula in the remainder of this paper. As a simple example, we consider the 2nd R\'enyi entropy firstly. These coordinates on UHP can be expressed by the original spacetime coordinate as follows.
\begin{eqnarray}\label{Transformationpoint}
\xi_1=-i\frac{(\frac{\omega_1+l}{\omega_1-l})^{1/2}+1}{(\frac{\omega_1+l}{\omega_1-l})^{1/2}-1},&&
\xi_2=-i\frac{(\frac{\omega_2+l}{\omega_2-l})^{1/2}+1}{(\frac{\omega_2+l}{\omega_2-l})^{1/2}-1},\nonumber \\
\xi_3=-i\frac{-(\frac{\omega_1+l}{\omega_1-l})^{1/2}+1}{-(\frac{\omega_1+l}{\omega_1-l})^{1/2}-1},&&
\xi_4=-i\frac{-(\frac{\omega_2+l}{\omega_2-l})^{1/2}+1}{-(\frac{\omega_2+l}{\omega_2-l})^{1/2}-1},\nonumber \\
\bar \xi_1=i\frac{(\frac{\bar \omega_1+l}{\bar \omega_1-l})^{1/2}+1}{(\frac{\bar \omega_1+l}{\bar \omega_1-l})^{1/2}-1},&&
\bar \xi_2=i\frac{(\frac{\bar \omega_1+l}{\bar \omega_1-l})^{1/2}+1}{(\frac{\bar \omega_1+l}{\bar \omega_1-l})^{1/2}-1},\nonumber \\
\bar \xi_3=i\frac{-(\frac{\bar \omega_1+l}{\bar \omega_1-l})^{1/2}+1}{-(\frac{\bar \omega_1+l}{\bar \omega_1-l})^{1/2}-1},&&
\bar \xi_4=i\frac{-(\frac{\bar \omega_1+l}{\bar \omega_1-l})^{1/2}+1}{-(\frac{\bar \omega_1+l}{\bar \omega_1-l})^{1/2}-1}.
\end{eqnarray}
One can analyze the above formula carefully in two different time regions.
When $0<t<L-l$ or $t>l+L$ with $\epsilon \rightarrow 0$,
\begin{eqnarray}\label{Varianceofpoint}
\xi_1-\xi_2 \simeq \frac{2\xi_1\epsilon}{\sqrt{\omega_1+l}\sqrt{\omega_1-l}},\nonumber \\
\xi_3-\xi_4 \simeq \frac{2\xi_3\epsilon}{\sqrt{\omega_1+l}\sqrt{\omega_1-l}},\nonumber \\
\bar \xi_1-\bar \xi_2 \simeq \frac{2\bar \xi_1\epsilon}{\sqrt{\bar \omega_1-l}\sqrt{\bar \omega_1+l}},\nonumber \\
\bar \xi_3-\bar \xi_4 \simeq \frac{2\bar \xi_3\epsilon}{\sqrt{\bar \omega_1-l}\sqrt{\bar \omega_1+l}},
\end{eqnarray}
and the Jacobi factor of conformal transformation,
\begin{eqnarray}
\frac{d\xi_1}{d\omega_1}&\simeq& \frac{d\xi_2}{d\omega_2}\simeq{1\over 2 i\epsilon}\xi_{1 2},\text{ }\text{}\frac{d\bar{\xi}_1}{d\bar{\omega}_1}\simeq \frac{d\bar{\xi}_2}{d\bar{\omega}_2}\simeq -{1\over 2 i\epsilon}\bar{\xi}_{1 2}\label{DerivitiveGeneralzationfour12}\\
\frac{d\xi_3}{d\omega_3}&\simeq& \frac{d\xi_4}{d\omega_4}\simeq {1\over 2 i\epsilon}\xi_{34}, \text{ }\text{ }\frac{d\bar{\xi}_3}{d\bar{\omega}_3}\simeq \frac{d\bar{\xi}_4}{d\bar{\omega}_4}\simeq -{1\over 2 i\epsilon}\bar{\xi}_{34}\label{DerivitiveGeneralzationfour34}\\
\end{eqnarray}
When $L-l<t<L+l$ with $\epsilon \rightarrow 0$,
\begin{eqnarray}\label{Varianceofpoitmiddletime}
\xi_4-\xi_1 \simeq \frac{2\xi_1\epsilon}{\sqrt{\omega_1+l}\sqrt{\omega_1-l}},\nonumber \\
\xi_2-\xi_3 \simeq \frac{2\xi_2\epsilon}{\sqrt{\omega_1+l}\sqrt{\omega_1-l}},\nonumber \\
\bar \xi_1-\bar \xi_2 \simeq -\frac{2\bar \xi_1\epsilon}{\sqrt{\bar \omega_1-l}\sqrt{\bar \omega_1+l}},\nonumber \\
\bar \xi_3-\bar \xi_4 \simeq -\frac{2\bar \xi_3\epsilon}{\sqrt{\bar \omega_1-l}\sqrt{\bar \omega_1+l}},
\end{eqnarray}
and the Jacobi factor of conformal transformation,
\begin{eqnarray}
\frac{d{\xi}_1}{d{\omega}_1}&\simeq& \frac{d{\xi}_4}{d{\omega}_4}={1\over 2 i\epsilon}{\xi}_{1 4}, \text{ }\text{ }\frac{d\bar{\xi}_1}{d\bar{\omega}_1}\simeq \frac{d\bar{\xi}_2}{d\bar{\omega}_2}\simeq-{1\over 2 i\epsilon}\bar{\xi}_{1 2}\label{DerivitiveGeneralzationfour21}\\
\frac{d\xi_2}{d\omega_2}&\simeq& \frac{d\xi_3}{d\omega_3}\simeq{1\over 2 i\epsilon}\xi_{23},\text{ }\text{ }\frac{d\bar{\xi}_3}{d\bar{\omega}_3}\simeq \frac{d\bar{\xi}_4}{d\bar{\omega}_4}\simeq -{1\over 2 i\epsilon}\bar{\xi}_{34}\label{DerivitiveGeneralzationfour22}
\end{eqnarray}
\subsection{2nd R\'enyi entropy for free boson}
 We will focus on the  following local operators in the free scalar field firstly,
\begin{eqnarray}\label{OperatorBoson}
O_1=e^{i\phi/2}, \ \ O_2=\frac{1}{\sqrt{2}}(e^{i\phi/2}+e^{-i\phi/2}).
\end{eqnarray}
The time evolution of R\'enyi entropy for such operators have already been studied in \cite{He:2014mwa} in 2D CFT on the complex plane. There are two kinds of boundary conditions for 2D free scalar field theory. One is $\frac{\partial \phi}{\partial n}|_B=0$ called Neumann boundary condition, the other is $\phi|_B=0$ called Dirichlet boundary condition. Since the boundary condition is homogenous, it is invariant under the conformal transformation.\\
 The image method \cite{CFT} \footnote{In conformal field theory literature, the image method is also called double trick sometimes.} is an efficient way to obtain the correlation function on UHP from correlation function on the full complex plane. The two kinds of boundary conditions correspond to different parity transformation in image method. Due to the presence of boundary, there are constraints on local conformal transformation, the anti-holomorphic and the holomorphic sectors in correlation function are no longer independent. More precisely, the correlation function on the upper half plane can be expressed by
holomorphic part of conformal block on the full complex plane with including the `images' of the holomorphic coordinates. That is to say,
\begin{eqnarray}
\langle \phi(z_1,\bar z_1)\phi(z_2,\bar z_2)...\phi(z_n,\bar z_n)\rangle_{UHP}
\end{eqnarray}
equals to
\begin{eqnarray}\label{4ninR2}
\langle \phi(z_1)\bar \phi( {z_1}^*)\phi(z_2)\bar \phi( {z_2}^*)...\phi(z_n)\bar \phi(z_n^*)\rangle_{R^2},
\end{eqnarray}
where $\phi$ and $\bar \phi$ refer to the holomorphic and anti-holomorphic part of the field $\phi$
{\color{red}\footnote{For most primary fields, it is impossible to divide the field $\phi$ into holomorphic and anti-holomorphic part. Here we only want to express that the $n$-point correlation functions on UHP which are dependent on the holomorphic conformal blocks of $2n$-point correlation functions on the full complex plane. In 2D free field theory, the conformal blocks of the operator $O_1$ is trivial. In this paper, we will also show the image method in 2D Ising model and other generic 2D CFTs.}}. After a parity transformation the anti-holomorphic part become a holomorphic field with conformal dimension of the original anti-holomorphic part. In terms of image method, we should introduce parity transformation. For the free boson the parity transformation is
\begin{eqnarray}\label{ParityBoson}
\phi(z,\bar z)=\eta \phi(\bar z,z),\ \ \eta=\pm 1,
\end{eqnarray}
$\eta=1,-1$ corresponds to the Neumann boundary condition and Dirichlet boundary condition respectively.
In terms of (\ref{result1}), to obtain R\'enyi entropy, it is necessary to know the two-point and four-point correlation function on the UHP.
\subsubsection{Local excitation $O_1$}\label{3.2.1}
Let's consider the operator $O_1$ firstly. Using the image method, we could get the two-point function
\begin{eqnarray}\label{twopointfunction}
\langle O_1^{\dag}(\omega_1,\bar \omega_1)O_1(\omega_2,\bar \omega_2)\rangle_{B_1}&=&\prod_{i=1}^2(\frac{d\omega_i}{d\xi_i^{'}})^{-h}(\frac{d\bar \omega_i}
{d\bar \xi_i^{'}})^{-\bar h}\langle O_1^{\dag}(\xi_1^{'},\bar \xi_1^{'})O_1(\xi_2^{'},\bar \xi_2^{'})\rangle_{UHP}\nonumber \\&=&\prod_{i=1}^2(\frac{d\omega_i}{d\xi_i^{'}})^{-h}(\frac{d\bar \omega_i}{d\bar \xi_i^{'}})^{-\bar h}\Big[\langle O_1^{\dag}(\xi_1^{'},\bar \xi_1^{'})O_1(\xi_2^{'},\bar \xi_2^{'})\tilde{O_1}^\dag (\xi_3^{'},\bar \xi_3^{'})\tilde{O_1}(\xi_4^{'},\bar \xi_4^{'})\rangle_{R^2}\Big]_{\text{holo}}\nonumber \\
&\simeq &\frac{1}{(\xi^{'}_{12}\xi^{'}_{34})^{1/4}}=\frac{1}{(4\epsilon^2)^{1/4}}.
\end{eqnarray}
where $\xi_3^{'}\equiv \xi_1^{'*}$,$\xi_4\equiv \xi_2^{'*}$,$\xi_i^{'}=i\omega_i$, $h=\bar h=1/8$ and $\tilde{O_1}$ is the field with parity transformation. The subindex `holo' means that we only keep the holomorphic part of the correlation and set the anti-holomorphic part to be constant  \footnote{We can use this rule in 2D free scalar field theory, since the conformal blocks related $O_1$ are actually trivial. We would like to appreciate communication with Cardy on this point.} which is determined by boundary condition in general. For two-point function one could normalize the field and take constant to be $1$.
The 4-point correlation function could be obtained by similar procedure.
\begin{eqnarray}\label{Fourpoitfunction}
&&\langle O_1^{\dag}(\omega_1,\bar \omega_1)O_1(\omega_2,\bar \omega_2)O_1^{\dag}(\omega_3,\bar \omega_3)O_1(\omega_4,\bar \omega_4)\rangle_{B_2}\nonumber \\
&&=\prod_{i=1}^4(\frac{d\omega_i}{d\xi_i})^{-h}(\frac{d\bar \omega_i}{d\bar \xi_i})^{-\bar h}\langle O_1^{\dag}(\xi_1,\bar \xi_1)O_1(\xi_2,\bar \xi_2)O_1^{\dag}(\xi_3,\bar \xi_3)O_1(\xi_4,\bar \xi_4)\rangle_{UHP}\nonumber \\
&&= \prod_{i=1}^4(\frac{d\omega_i}{d\xi_i})^{-h}(\frac{d\bar \omega_i}{d\bar \xi_i})^{-\bar h}\Big[\langle O_1^{\dag}(\xi_1,\bar \xi_1)...O_1(\xi_4,\bar \xi_4)\tilde{O_1}^\dag (\xi_5,\bar \xi_5)...\tilde{O_1}(\xi_8,\bar \xi_8)\rangle_{R^2}\Big]_{\text{holo}},\nonumber \\
&&
\end{eqnarray}
where $\xi_5=\bar \xi_1$, $\xi_6=\bar \xi_2$, $\xi_7=\bar \xi_3$, $\xi_8=\bar \xi_4$, the operator $\tilde{O}_1=e^{i\eta \phi}$ on $R^2$ with parity transformation.

In the region $0<t<L-l$ or $t>l+L$, as (\ref{Varianceofpoint}) shows, the correlation function (\ref{Fourpoitfunction}) can be factorized as
\begin{eqnarray}\label{factorized}
&&\langle O_1^{\dag}(\xi_1,\bar \xi_1)...O_1(\xi_4,\bar \xi_4)\tilde{O_1}^\dag (\xi_5,\bar \xi_5)...\tilde{O_1}(\xi_8,\bar \xi_8)\rangle_{R^2}\nonumber \\
&&\propto\langle O_1^{\dag}(\xi_1,\bar \xi_1)O_1(\xi_2,\bar \xi_2)\rangle_{R^2}...\langle \tilde{O_1}^{\dag}(\xi_7,\bar \xi_7)\tilde{O_1}(\xi_8,\bar \xi_8)\rangle_{R^2}
\end{eqnarray} in $\epsilon\rightarrow 0$ limit. The image method leaves us with a constant C. To fix the constant C, we take the limit $\xi_1\to \xi_2$, $\bar \xi_1 \to \bar \xi_2$, $\xi_3\to \xi_4$ and $\bar \xi_3 \to \bar \xi_4$ in (\ref{Fourpoitfunction}), one could find
\begin{eqnarray}
&&\langle O_1^{\dag}(\xi_1,\bar \xi_1)O_1(\xi_2,\bar \xi_2)O_1^{\dag}(\xi_3,\bar \xi_3)O_1(\xi_4,\bar \xi_4)\rangle_{UHP}\nonumber \\
&\simeq& \langle O_1^{\dag}(\xi_1,\bar \xi_1)O_1(\xi_2,\bar \xi_2)\rangle_{UHP} \langle O_1^{\dag}(\xi_3,\bar \xi_3)O_1(\xi_4,\bar \xi_4)\rangle_{UHP}.\label{blockCC}
\end{eqnarray}
In terms of image method, we can also obtain
\begin{eqnarray}
&&\langle O_1^{\dag}(\xi_1,\bar \xi_1)O_1(\xi_2,\bar \xi_2)O_1^{\dag}(\xi_3,\bar \xi_3)O_1(\xi_4,\bar \xi_4)\rangle_{UHP}\nonumber \\
&=&\Big[\langle O_1^{\dag}(\xi_1,\bar \xi_1)...O_1(\xi_4,\bar \xi_4)O(\xi_2,\bar \xi_2)\tilde{O_1}^\dag (\xi_5,\bar \xi_5)...\tilde{O_1}(\xi_8,\bar \xi_8)\rangle_{R^2}\Big]_{\text{holo}}\nonumber\\
&\simeq& \frac{C }{(\xi_{12}\xi_{34}\xi_{56}\xi_{78})^{1/4}},\label{blockC}
\end{eqnarray}
where $C$ is constant. Comparing (\ref{blockCC}) with (\ref{blockC}), we can fix $C=1$ which is consistent with the normalization of the two point correlation function.

For $n=2$, the variation of the R\'enyi entropy (\ref{result1}) is
\begin{eqnarray}\label{O1result1}
\Delta S^{(2)}_A&=&-\log \Big[\prod_{k=1}^{4} (\frac{d\omega_k}{d\xi_k})^{-h}(\frac{d\omega_k}{d\xi_k})^{-\bar h}\frac{\langle O^{\dag}(\xi_1,\bar \xi_1)O(\xi_2,\bar \xi_2)O^{\dag}(\xi_{3},\bar \xi_{3})O(\xi_{4},\bar \xi_{4})\rangle_{UHP}}{(\langle O^{\dag}(\omega_1,\bar \omega_1)O(\omega_2,\bar \omega_2)\rangle_{B_1})^2}\Big]\nonumber\\
&=& -\log \Big[(\frac{(\xi_{12}\xi_{34}\xi_{56}\xi_{78})^{2}}{(2\epsilon)^{8}})^{1/8}\frac{(4\epsilon^2)^{1/2}}{(\xi_{12}\xi_{34}\xi_{56}\xi_{78})^{1/4}}\Big]=0,
\end{eqnarray}
where we have used (\ref{twopointfunction}), (\ref{Fourpoitfunction}),(\ref{blockC}) and Jacobi factor (\ref{DerivitiveGeneralzationfour12})(\ref{DerivitiveGeneralzationfour34}).

In the other region $L-l<t<L+l$, we can not factorize the correlation function as (\ref{factorized}) directly.
For 2D free scalar theory, the situation becomes much simpler. The correlation function could be expressed as \cite{CFT}
\begin{eqnarray}\label{Bosoncorrelation}
\langle e^{i\alpha_1\phi}...e^{i\alpha_n \phi}\rangle=\prod_{i<j}[z_{ij}^{\alpha_i\alpha_j}][\bar z_{ij}^{\alpha_i\alpha_j}],
\end{eqnarray}
with the neutral condition $\alpha_1+\alpha_2+...\alpha_n=0$  and $z_{ij}=z_i-z_j, \bar{z}_{ij}=\bar{z}_i-\bar{z}_j $.
Thus
\begin{eqnarray}\label{8pointfunction}
&&\langle O_1^{\dag}(\xi_1,\bar \xi_1)...O_1(\xi_4,\bar \xi_4)\rangle_{UHP}
\nonumber\\
&=&\Big[\langle O_1^{\dag}(\xi_1,\bar \xi_1)...O_1(\xi_4,\bar \xi_4)O(\xi_2,\bar \xi_2)\tilde{O_1}^\dag (\xi_5,\bar \xi_5)...\tilde{O_1}(\xi_8,\bar \xi_8)\rangle_{R^2}\Big]_{\text{holo}}\nonumber\\
&\simeq& \frac{1}{(\xi_{41}\xi_{23}\xi_{56}\xi_{78})^{1/4}},
\end{eqnarray}
where we have fix  the constant to be 1.
To get $\Delta S^{(2)}_A$ we only need to change $2\leftrightarrow 4$ in (\ref{O1result1}). Using (\ref{Varianceofpoitmiddletime})(\ref{twopointfunction})(\ref{8pointfunction}) and Jacobi factor (\ref{DerivitiveGeneralzationfour21})(\ref{DerivitiveGeneralzationfour22}), we get
\begin{eqnarray}
\Delta S^{(2)}_A= -\log \Big[(\frac{(\xi_{14}\xi_{32}\xi_{56}\xi_{78})^{2}}{(2\epsilon)^{8}})^{1/8}\frac{(4\epsilon^2)^{1/2}}{(\xi_{14}\xi_{32}\xi_{56}\xi_{78})^{1/4}}\Big]=0,
\end{eqnarray}
in $L-l<t<L+l$.
To close this subsection, one more thing should be noted that $\Delta S^{(2)}_A$ does not depend on the choice of parity transformation. Actually in our above calculation we do not use the value of $\eta$, which is related to the boundary condition. The leading order of the correlation function is same. $\Delta S^{(2)}_A$ is always zero for operator $O_1$. To see the effect of the boundary, we should consider more complicated example.
\subsubsection{Local excitation $O_2$}
$O_2$ is a linear combination between $O_1$ and $O^{\dag}_1$. We can calculate $\Delta S^{(2)}_A$ with following the logic in previous section.  The two-point correlation function of $O_2$,
\begin{eqnarray}\label{2pointfunctionO2}
&&\langle O^{\dag}_2(\xi_1^{'},\bar \xi_1^{'}) O_2(\xi_2^{'},\bar \xi_2^{'})\rangle_{UHP}= \frac{1}{2}\Big[\langle O^{\dag}_1(\xi_1^{'},\bar \xi_1^{'}) O^{\dag}_1(\xi_2^{'},\bar \xi_2^{'})\rangle_{UHP} +\langle O_1(\xi_1^{'},\bar
\xi_1^{'})O_1(\xi_2^{'},\bar \xi_2^{'})\rangle_{UHP} \nonumber \\
&&+\langle O_1(\xi_1^{'},\bar \xi_1^{'}) O^{\dag}_1(\xi_2^{'},\bar \xi_2^{'})\rangle_{UHP}+\langle O^{\dag}_1(\xi_1^{'},\bar
\xi_1^{'}) O_1(\xi_2^{'},\bar \xi_2^{'})\rangle_{UHP} \Big]\nonumber \\
&\simeq& \frac{1}{(\xi_{12}^{'})^{1/4}(\xi_{34}^{'})^{1/4}}= \frac{1}{(4\epsilon^2)^{1/4}},
\end{eqnarray}
in the limit $\epsilon \to 0$. Here we have used the parity transformation (\ref{ParityBoson}) related to the boundary condition. We have set anti-holomorphic parts to be $1$.

The four-point correlation function on UHP is
\begin{eqnarray}\label{O2operator4pointfunction}
\langle O_2^{\dag}(\xi_1,\bar \xi_1)...O_2(\xi_4,\bar \xi_4)\rangle_{UHP}
&&=\frac{1}{4}\Big[\langle O_1(\xi_1,\bar \xi_1)O^{\dag}_1(\xi_2,\bar \xi_2) O_1(\xi_3,\bar \xi_3)O^{\dag}_1(\xi_4,\bar \xi_4) \rangle_{UHP}\nonumber \\
&&+\langle O_1(\xi_1,\bar \xi_1)O_1(\xi_2,\bar \xi_2) O^{\dag}_1(\xi_3,\bar \xi_3)O^{\dag}_1(\xi_4,\bar \xi_4) \rangle_{UHP}\nonumber \\
&&+\langle O^{\dag}_1(\xi_1,\bar \xi_1)O_1(\xi_2,\bar \xi_2) O_1(\xi_3,\bar \xi_3)O^{\dag}_1(\xi_4,\bar \xi_4) \rangle_{UHP}\nonumber \\
&&+\langle O_1(\xi_1,\bar \xi_1)O^{\dag}_1(\xi_2,\bar \xi_2) O^{\dag}_1(\xi_3,\bar \xi_3)O_1(\xi_4,\bar \xi_4) \rangle_{UHP}\nonumber \\
&&+\langle O^{\dag}_1(\xi_1,\bar \xi_1)O_1(\xi_2,\bar \xi_2) O^{\dag}_1(\xi_3,\bar \xi_3)O_1(\xi_4,\bar \xi_4) \rangle_{UHP}\nonumber \\
&&+\langle O^{\dag}_1(\xi_1,\bar \xi_1)O^{\dag}_1(\xi_2,\bar \xi_2) O_1(\xi_3,\bar \xi_3)O_1(\xi_4,\bar \xi_4) \rangle_{UHP}\nonumber \\
&&+\langle O^{\dag}_1(\xi_1,\bar \xi_1)O^{\dag}_1(\xi_2,\bar \xi_2) O^{\dag}_1(\xi_3,\bar \xi_3)O^{\dag}_1(\xi_4,\bar \xi_4) \rangle_{UHP}\nonumber \\
&&+\langle O_1(\xi_1,\bar \xi_1)O^{\dag}_1(\xi_2,\bar \xi_2) O^{\dag}_1(\xi_3,\bar \xi_3)O^{\dag}_1(\xi_4,\bar \xi_4) \rangle_{UHP}\nonumber \\
&&+\langle O^{\dag}_1(\xi_1,\bar \xi_1)O^{\dag}_1(\xi_2,\bar \xi_2) O_1(\xi_3,\bar \xi_3)O^{\dag}_1(\xi_4,\bar \xi_4) \rangle_{UHP}\nonumber \\
&&+\langle O^{\dag}_1(\xi_1,\bar \xi_1)O_1(\xi_2,\bar \xi_2) O^{\dag}_1(\xi_3,\bar \xi_3)O^{\dag}_1(\xi_4,\bar \xi_4) \rangle_{UHP}\nonumber \\
&&+\langle O_1(\xi_1,\bar \xi_1)O_1(\xi_2,\bar \xi_2) O_1(\xi_3,\bar \xi_3)O^{\dag}_1(\xi_4,\bar \xi_4) \rangle_{UHP} \nonumber \\
&&+\langle O^{\dag}_1(\xi_1,\bar \xi_1)O^{\dag}_1(\xi_2,\bar \xi_2) O^{\dag}_1(\xi_3,\bar \xi_3)O_1(\xi_4,\bar \xi_4) \rangle_{UHP}\nonumber \\
&&+\langle O_1(\xi_1,\bar \xi_1)O_1(\xi_2,\bar \xi_2) O^{\dag}_1(\xi_3,\bar \xi_3)O_1(\xi_4,\bar \xi_4) \rangle_{UHP}\nonumber \\
&&+\langle O_1(\xi_1,\bar \xi_1)O^{\dag}_1(\xi_2,\bar \xi_2) O_1(\xi_3,\bar \xi_3)O_1(\xi_4,\bar \xi_4) \rangle_{UHP} \nonumber \\
&&+\langle O^{\dag}_1(\xi_1,\bar \xi_1)O_1(\xi_2,\bar \xi_2) O_1(\xi_3,\bar \xi_3)O_1(\xi_4,\bar \xi_4) \rangle_{UHP}\nonumber \\
&&+\langle O_1(\xi_1,\bar \xi_1)O_1(\xi_2,\bar \xi_2) O_1(\xi_3,\bar \xi_3)O_1(\xi_4,\bar \xi_4) \rangle_{UHP}\Big].
\end{eqnarray}
We could use the result of four-point function of $O_1$ which has been studied in section (\ref{3.2.1}).
From (\ref{ParityBoson})and (\ref{4ninR2}) one could see that the four-point function is dependent on the boundary conditions or parity transformations. We will calculate $\Delta S^{(2)}_A$ with Neumann and Dirichlet boundary condition respectively.

1.\  For the Neumann boundary condition, i.e., $\eta=1$.

In (\ref{O2operator4pointfunction}), terms containing equal number of $O_1$ and $O^{\dag}_1$ will survive due to the neutrality condition. Thus there are 6 terms making contribution to the 4-point correlation function. In the region $0<t<L-l$ or $t>L+l$, (\ref{O2operator4pointfunction}) can be expressed by factorized form on the $R^2$. For example, the first term in (\ref{O2operator4pointfunction}) as a leading term is
\begin{eqnarray}
&&\langle O_1(\xi_1,\bar \xi_1)O^{\dag}_1(\xi_2,\bar \xi_2) O_1(\xi_3,\bar \xi_3)O^{\dag}_1(\xi_4,\bar \xi_4) \rangle_{UHP}\nonumber \\
&=&\Big[\langle O_1(\xi_1,\bar \xi_1)O^{\dag}_1(\xi_2,\bar \xi_2) O_1(\xi_3,\bar \xi_3)O^{\dag}_1(\xi_4,\bar \xi_4)O_1(\xi_5,\bar \xi_5) O^{\dag}_1(\xi_6,\bar \xi_6)O_1(\xi_7,\bar \xi_7) O^{\dag}_1(\xi_8,\bar \xi_8)\rangle\Big]_{\text{holo}}\nonumber \\
&\simeq& \Big[\langle O_1(\xi_1,\bar \xi_1)O^{\dag}_1(\xi_2,\bar \xi_2)\rangle \langle O_1(\xi_3,\bar \xi_3)O^{\dag}_1(\xi_4,\bar \xi_4) \rangle \langle O_1(\xi_5,\bar \xi_5)O_1(\xi_6,\bar \xi_6)\rangle\nonumber \\
 &&\langle O^{\dag}_1(\xi_7,\bar \xi_7)O^{\dag}_1(\xi_8,\bar \xi_8)\rangle\Big]_{\text{holo}}=\frac{1}{(\xi_{12}\xi_{34}\xi_{56}\xi_{78})^{1/4}}\sim \frac{1}{\epsilon},
\end{eqnarray}
in the limit $\epsilon \to 0$.
The second terms as sub-leading term is
\begin{eqnarray}
&&\langle O_1(\xi_1,\bar \xi_1)O_1(\xi_2,\bar \xi_2) O^{\dag}_1(\xi_3,\bar \xi_3)O^{\dag}_1(\xi_4,\bar \xi_4)\rangle_{UHP}=\Big[\langle O_1(\xi_1,\bar \xi_1)O_1(\xi_2,\bar \xi_2) \nonumber \\
&&O^{\dag}_1(\xi_3,\bar \xi_3)O^{\dag}_1(\xi_4,\bar \xi_4)O_1(\xi_5,\bar \xi_5)O_1(\xi_6,\bar \xi_6) O^{\dag}_1(\xi_7,\bar \xi_7)O^{\dag}_1(\xi_8,\bar \xi_8)\rangle_{R_2}\Big]_{\text{holo}} \nonumber \\
&\simeq& (\frac{\xi_{12}\xi_{34}...\xi_{56}\xi_{78}}{\xi_{13}\xi_{14}...\xi_{57}\xi_{58}})^{1/4}\sim \epsilon,
\end{eqnarray}
in the limit $\epsilon \to 0$. Totally, there are four terms that are of order $O(\epsilon^{-1})$ in (\ref{O2operator4pointfunction}).

Using (\ref{result1}), we get $\Delta S_A^{(n)}$ for $n=2$,
\begin{eqnarray}
\Delta S^{(2)}_A&\simeq &-\log \Big[(\frac{(\xi_{12}\xi_{34}\xi_{56}\xi_{78})^{2}}{(2\epsilon)^{8}})^{1/8} \frac{\langle O_2^{\dag}(\xi_1,\bar \xi_1)...O_2(\xi_4,\bar
\xi_4)\rangle_{UHP}}{(\langle O^{\dag}_2(\xi_1,\bar \xi_1) O_2(\xi_2,\bar \xi_2)\rangle_{UHP})^2}\Big]\nonumber \\
&=&-\log \Big[(\frac{(\xi_{12}\xi_{34}\xi_{56}\xi_{78})^{2}}{(2\epsilon)^{8}})^{1/8}\frac{(4\epsilon^2)^{1/2}}{(\xi_{12}\xi_{34}\xi_{56}\xi_{78})^{1/4}}\Big]=0,
\end{eqnarray}
In $L-l<t<L+l$ with $\epsilon \to 0$, (\ref{Varianceofpoitmiddletime}) shows $\xi_{14}\sim \xi_{23}\sim \epsilon$. Terms making non-vanishing contribution to 4-point correlation function will change in the limit $\epsilon \to 0$. For example the third term in (\ref{O2operator4pointfunction}) as a sub-leading term is,
\begin{eqnarray}
&&\langle O^{\dag}_1(\xi_1,\bar \xi_1)O_1(\xi_2,\bar \xi_2) O_1(\xi_3,\bar \xi_3)O^{\dag}_1(\xi_4,\bar \xi_4) \rangle_{UHP}\nonumber \\
&=&\Big[\langle O^{\dag}_1(\xi_1,\bar \xi_1)O_1(\xi_2,\bar \xi_2) O_1(\xi_3,\bar \xi_3)O^{\dag}_1(\xi_4,\bar \xi_4)O^{\dag}_1(\xi_5,\bar \xi_5)O_1(\xi_6,\bar \xi_6) O_1(\xi_7,\bar \xi_7)O^{\dag}_1(\xi_8,\bar \xi_8) \rangle_{R^2}\Big]_{\text{holo}}\nonumber \\
&=&(\frac{\xi_{14}\xi_{23}...\xi_{58}\xi_{67}}{\xi_{12}\xi_{34}...\xi_{56}\xi_{78}})^{1/4}\sim O(1),
\end{eqnarray}
where `...' stands for the terms that are $O(1)$ in the limit $\epsilon \to 0$. One could count the leading contribution term by term in (\ref{O2operator4pointfunction}), there are only two terms that are of $O(\epsilon^{-1})$. Thus the variation of the R\'enyi entropy $\Delta S^{(n)}_{A}$ for $n=2$ is
\begin{eqnarray}
\Delta S^{(2)}_A&\simeq &-\log \Big[(\frac{(\xi_{14}\xi_{32}\xi_{56}\xi_{78})^{2}}{(2\epsilon)^{8}})^{1/8} \frac{\langle O_2^{\dag}(\xi_1,\bar \xi_1)...O_2(\xi_4,\bar
\xi_4)\rangle_{UHP}}{(\langle O^{\dag}_2(\xi_1,\bar \xi_1) O_2(\xi_2,\bar \xi_2)\rangle_{UHP})^2}\Big]\nonumber \\
&=&-\log \Big[(\frac{(\xi_{14}\xi_{32}\xi_{56}\xi_{78})^{2}}{(2\epsilon)^{8}})^{1/8}\frac{(4\epsilon^2)^{1/2}}{2(\xi_{14}\xi_{23}\xi_{56}\xi_{78})^{1/4}}\Big]=\log 2,
\end{eqnarray}

2.\ \ For the Dirichlet boundary condition, i.e., $\eta =-1$.\\
All the terms in (\ref{O2operator4pointfunction}) are non-vanishing in term of the neutrality condition. However, there are still four different terms in (\ref{O2operator4pointfunction}) contributing to the leading order in the region $L+l<t$ or $t<L-l$ in $\epsilon \rightarrow 0$ limit. Thus the ratio (\ref{result1}) is 1 and $\Delta S^{(2)}_A=0$. In the region $L-l<t<L+l$, there are {\color{black}two} terms in (\ref{O2operator4pointfunction}) with $\epsilon \rightarrow 0$ limit, which is the same as the situation of Neumann boundary condition. Thus the ratio (\ref{result1}) is ${1\over 2}$ and $\Delta S^{(2)}_A=\log 2$. In this example we could see that $\Delta S^{(2)}_A$ does not depend on the choice of boundary conditions.

\subsection{n-th R\'enyi Entropy for free boson}
In this subsection, we would like to generalize our studies to n-th R\'enyi entropy which involves in the the 2n-point correlation function on $B_n$.
The conformal transformation (\ref{Transformation1}) (\ref{Transformation2}) can map $B_n$ to UHP. Finally one can calculate the 2n-point correlation function on UHP by the `method of images' in terms of 4n-point correlation function on $R_2$. The points $\xi_1,\xi_2...\xi_{2n}$ on $B_n$ are
\begin{eqnarray}\label{Npointaftermapping}
&&\xi_{2k+1}=-i \frac{e^{2i\pi k/n}(\frac{\omega_1+l}{\omega_1-l})^{1/n}+1}{e^{2i\pi k/n}(\frac{\omega_1+l}{\omega_1-l})^{1/n}-1},\ \ \ \
\xi_{2k+2}=-i \frac{e^{2i\pi k/n}(\frac{\omega_2+l}{\omega_2-l})^{1/n}+1}{e^{2i\pi k/n}(\frac{\omega_2+l}{\omega_2-l})^{1/n}-1},\nonumber \\
&&\bar \xi_{2k+1}=-i \frac{e^{2i\pi k/n}(\frac{\bar \omega_1+l}{\bar \omega_1-l})^{1/n}+1}{e^{2i\pi k/n}(\frac{\bar \omega_1+l}{\bar \omega_1-l})^{1/n}-1},\ \ \ \
\bar \xi_{2k+2}=-i \frac{e^{2i\pi k/n}(\frac{\bar \omega_2+l}{\bar \omega_2-l})^{1/n}+1}{e^{2i\pi k/n}(\frac{\bar \omega_2+l}{\bar \omega_2-l})^{1/n}-1},
\end{eqnarray} where $0\le k\le n-1$.
In the region $t>L+l$ or $t<L-l$, one can obtain
\begin{eqnarray}\label{varianceofpoint1}
&&\xi_{2k+1}-\xi_{2k+2}\simeq \frac{8ie^{2i\pi k/n} l \epsilon}{n(\omega_1-l)^{1-1/n}(\omega_1+l)^{1-1/n}[e^{2i\pi k/n}(\omega_1-l)^{1/n}-(\omega_1+l)^{1/n}]^2}\nonumber \\
&&\bar \xi_{2k+1}-\bar \xi_{2k+2}\simeq \frac{-8ie^{2i\pi k/n} l \epsilon}{n(\bar \omega_1-l)^{1-1/n}(\bar \omega_1+l)^{1-1/n}[e^{2i\pi k/n}(\bar \omega_1-l)^{1/n}-(\bar \omega_1+l)^{1/n}]},\nonumber \\
\
\end{eqnarray}
and the Jacobi factor of conformal transformation,
\begin{eqnarray}
\frac{d\xi_{2k+1}}{d\omega_{2k+1}}&\simeq&\frac{d\xi_{2k+2}}{d\omega_{2k+2}}\simeq{1\over 2 i\epsilon}\xi_{2k+1, 2k+2}\label{DerivitiveGeneralzation11}\\
\frac{d\bar {\xi}_{2k+1}}{d\bar{\omega}_{2k+1}}&\simeq&\frac{d\bar {\xi}_{2k+2}}{d\bar{\omega}_{2k+2}}\simeq-{1\over 2 i\epsilon}\bar{\xi}_{2k+1, 2k+2}\label{DerivitiveGeneralzation12}
\end{eqnarray}
In $L-l<t<L+l$, one could find
\begin{eqnarray}\label{varianceofpoint2}
&&\xi_{2k}-\xi_{2k+1} \simeq \frac{8 i e^{2i\pi k/n} l  \epsilon }{n  (-l+\omega_1)^{1-1/n}(l+\omega_1)^{1-1/n}[e^{2i\pi k/n}(\omega_1-l)^{1/n}-(\omega_1+l)^{1/n}]^2}\nonumber \\
&&\bar \xi_{2k+1}-\bar \xi_{2k} \simeq \frac{-8ie^{2i\pi k/n} l \epsilon}{n(\bar \omega_1-l)^{1-1/n}(\bar \omega_1+l)^{1-1/n}[e^{2i\pi k/n}(\bar \omega_1-l)^{1/n}-(\bar \omega_1+l)^{1/n}]}
\end{eqnarray}
and the Jacobi factor of conformal transformation,
\begin{eqnarray}
\frac{d\xi_{2k}}{d\omega_{2k}}&\simeq&\frac{d\xi_{2k+1}}{d\omega_{2k+1}}\simeq{1\over 2 i\epsilon}\xi_{2k, 2k+1}\label{DerivitiveGeneralzation21}\\
\frac{d\bar{\xi}_{2k}}{d\bar{\omega}_{2k}}&\simeq&\frac{d\bar{\xi}_{2k+1}}{d\bar{\omega}_{2k+1}}\simeq-{1\over 2 i\epsilon}\bar{\xi}_{2k, 2k+1}\label{DerivitiveGeneralzation22}.
\end{eqnarray}
\subsubsection{Local excitation $O_1$}
We consider the operator $O_1$ firstly.
In the region $L-l<t$ or $t>L+l$ with $\epsilon \rightarrow 0$, the 2n-point correlation function of $O_1$ is
\begin{eqnarray}\label{BosonGeneraliztion2nO1}
&&\langle O_1^{\dag}(\xi_1,\bar \xi_1)O_1(\xi_2,\bar \xi_2)...O_1(\xi_{2n},\bar \xi_{2n})\rangle_{UHP}\nonumber \\
&&=\Big[\langle O_1^{\dag}(\xi_1,\bar \xi_1)...O_1(\xi_{2n},\bar \xi_{2n})\tilde{O}_1^{\dag}(\xi_{2n+1},\bar \xi_{2n+1})...\tilde{O}_1(\xi_{4n},\bar \xi_{4n})\rangle_{R_2}\Big]_{\text{holo}}\nonumber \\
&&\simeq \Big[\langle O_1^{\dag}(\xi_1,\bar \xi_1)O_1(\xi_{2},\bar \xi_{2})\rangle... \langle \tilde{O}_1^{\dag}(\xi_{4n-1},\bar \xi_{4n-1})\tilde{O}_1(\xi_{4n},\bar \xi_{4n})\rangle_{R_2}\Big]_{\text{holo}}\nonumber \\
&&=\frac{1}{(\xi_{12}...\xi_{2n-1,2n})^{1/4}(\xi_{2n+1,2n+2}...\xi_{4n-1,4n})^{1/4}},
\end{eqnarray}
where $\tilde{O}_1$ is defined as $e^{-i\eta \phi/2}$ with parity transformation and $\xi_{2n+1}=\bar \xi_{1}$,...,$\xi_{4n}=\bar \xi_{2n}$. Using (\ref{result1}), the variation of the $n$-th R\'enyi entropy can be obtained as follows.
\begin{eqnarray}
\Delta S^{(n)}_{A}&=&\frac{1}{1-n}\log \Big[\prod_{k=1}^{2n} (\frac{d\omega_k}{d\xi_k})^{-h}(\frac{d\omega_k}{d\xi_k})^{-\bar h}\frac{\langle O^{\dag}(\xi_1,\bar \xi_1)O(\xi_2,\bar \xi_2)...O^{\dag}(\xi_{2n-1},\bar \xi_{2n-1})O(\xi_{2n},\bar \xi_{2n})\rangle_{UHP}}{(\langle O^{\dag}(\omega_1,\bar \omega_1)O(\omega_2,\bar \omega_2)\rangle_{B_1})^n}\Big]\nonumber\\
&=&\frac{1}{1-n}\log \Big[ \Big(\frac{(\xi_{12}...\xi_{2n-1,2n})^{2}(\xi_{2n+1,2n+2}...\xi_{4n-1,4n})^{2}}{(2\epsilon)^{4n}}\Big)^{1/ 8} \nonumber\\&& \frac{(4\epsilon^2)^{{n/ 4}}}{(\xi_{12}...\xi_{2n-1,2n})^{1/4}(\xi_{2n+1,2n+2}...\xi_{4n-1,4n})^{1/4}}\Big]=0,
\end{eqnarray}
where we have used (\ref{twopointfunction})(\ref{varianceofpoint1})(\ref{BosonGeneraliztion2nO1}), and Jacobi factor (\ref{DerivitiveGeneralzation11})(\ref{DerivitiveGeneralzation12}).

In the region $L-l<t<L+l$ with $\epsilon\rightarrow 0$, the $2n$-point correlation function on UHP is
\begin{eqnarray}\label{BosonGeneralizationO1two}
&&\langle O_1^{\dag}(\xi_1,\bar \xi_1)O_1(\xi_2,\bar \xi_2)...O_1(\xi_{2n},\bar \xi_{2n})\rangle_{UHP}\nonumber \\
&\simeq& \frac{1}{(\xi_{23}...\xi_{2n,1})^{1/4}(\xi_{2n+1,2n+2}...\xi_{4n-1,4n})^{1/4}}.
\end{eqnarray}
Then, \begin{eqnarray}
\Delta S^{(n)}_{A}&=&\frac{1}{1-n}\log \Big[ \Big(\frac{(\xi_{23}...\xi_{2n,1})^{2}(\xi_{2n+1,2n+2}...\xi_{4n-1,4n})^{2}}{(2\epsilon)^{4n}}\Big)^{1/ 8} \nonumber\\&& \frac{(4\epsilon^2)^{{n/ 4}}}{(\xi_{23}...\xi_{2n,1})^{1/4}(\xi_{2n+1,2n+2}...\xi_{4n-1,4n})^{1/4}}\Big]=0,
\end{eqnarray}
where we have made use of (\ref{BosonGeneralizationO1two}) and Jacobi factor (\ref{DerivitiveGeneralzationfour21})(\ref{DerivitiveGeneralzationfour22}).
\subsubsection{Local excitation $O_2$}
For operator $O_2$, there are $2^{2n}$ terms making contribution to the correlation function like the ones in (\ref{O2operator4pointfunction}). Firstly let us consider the case with Neumman boundary condition. These terms with equal number of $O_1$ and $O^{\dag}_1$ can survive in the limit $\epsilon \rightarrow 0$. The $2n$-point correlation function of $O_1$ on $B_n$ can be expressed by $4n$-point correlation function on $R^2$,
\begin{eqnarray}\label{2npointfunction}
 &&\langle O_1^{\dag}(\xi_1,\bar \xi_1)O_1(\xi_2,\bar \xi_2)...O_1^{\dag}(\xi_{2n-1},\bar \xi_{2n-1})O_1(\xi_{2n},\bar \xi_{2n})\rangle_{UHP}\nonumber \\
 &=& \Big[\langle O_1^{\dag}(\xi_1,\bar \xi_1)...O_1(\xi_{2n},\bar \xi_{2n})\tilde{O}^{\dag}_1(\xi_{2n+1},\bar \xi_{2n+1})...\tilde{O}_1(\xi_{4n},\bar \xi_{4n})\rangle\Big]_{\text{holo}}.
\end{eqnarray}
 To  be convenient, we use the symbol $+1$ referring to $O^\dag_1$ and $-1$ referring to $O_1$ in the correlation function to simplify our analysis. Then the $2n$-point correlation function on UHP can be formally written as
 \begin{eqnarray}\label{Key2nBosonO22}
 &&\langle O_2^{\dag}(\xi_1,\bar \xi_1)O_2(\xi_2,\bar \xi_2)...O_2^{\dag}(\xi_{2n-1},\bar \xi_{2n-1})O_2(\xi_{2n},\bar \xi_{2n})\rangle_{UHP}\nonumber \\
 &=&\frac{1}{2^n}\sum_{\substack{i_1,i_2,...,i_{2n}=\pm1 \\ i_{2n+1}=i_1,...i_{4n}=i_{2n}} }\langle i_1,i_2,...,i_{2n},i_{2n+1}...,i_{4n} \rangle_{R^2},
 \end{eqnarray}
where we have made use of the image method  in the second line and $i_j=\pm 1$ stands for the operator $O_1$ or $O^{\dag}_1$ with the coordinate ($\xi_j$,$\bar \xi_j$). The constraints ${i_{2n+j}}={i_{j}}$ with $ 1\le j \le 2n$ corresponds to the Neumann boundary condition in terms of image method. For the correlation function in the Neumann boundary condition case, the non-zero terms in (\ref{Key2nBosonO22}) should satisfy neural condition
\begin{eqnarray}\label{Neumannconstraint}
i_1+i_2+...+i_{2n}+...i_{4n}=0.
\end{eqnarray}
In the region $t<L-l$ or $t>L+l$ with the limit $\epsilon\rightarrow 0$, due to (\ref{varianceofpoint1}), the leading contribution of the 2n-point correlation
function (\ref{Key2nBosonO22}) are
\begin{eqnarray}\label{free1st2n}
&&\sum_{\substack {i_1+i_2=0,i_3+i_4=0 \\...i_{4n-1}+i_{4n}=0 \\ {i_{2n+1}}={i_{1}}...{i_{4n}}={i_{2n}}}} \langle i_1,i_2,...,i_{2n},i_{2n+1}...,i_{4n} \rangle_{R^2}\nonumber \\
&&\simeq 2^n \langle i_1,i_2\rangle_{R^2}...\langle i_{2n-1},i_{2n}\rangle_{R^2} \langle i_{2n+1},i_{2n+2}\rangle_{R^2} ...\langle i_{4n-1},i_{4n} \rangle_{R^2},
\end{eqnarray}
 there are $2^n$ terms that are leading divergence after considering the constraints, and these terms are all equal to each other.
Equivalently, (\ref{Key2nBosonO22}) can be written by the notation $O_{1}$ and $O^{\dag}_1 $ as
\begin{eqnarray}
&&2^n \langle O_1^\dag(\xi_{1}) O_1(\xi_{2})\rangle_{R^2} ...\langle O_1^\dag(\xi_{2n-1}) O_1(\xi_{2n})\rangle_{R^2}...\langle O_1^\dag(\xi_{4n-1}) O_1(\xi_{4n})\rangle_{R^2}\nonumber \\
&=&\frac{2^n}{(\xi_{12}...\xi_{2n-1,2n})^{1/4}(\xi_{2n+1,2n+2}...\xi_{4n-1,4n})^{1/4}}.
\end{eqnarray}

Using (\ref{Key2nBosonO22}), the $2n$-point correlation function on $B_{2n}$ is
\begin{eqnarray}
&&\langle O_2^{\dag}(\xi_1,\bar \xi_1)O_2(\xi_2,\bar \xi_2)...O_2^{\dag}(\xi_{2n-1},\bar \xi_{2n-1})O_2(\xi_{2n},\bar \xi_{2n})\rangle_{UHP}\nonumber \\
&\simeq& \frac{1}{(\xi_{12}...\xi_{2n-1,2n})^{1/4}(\xi_{2n+1,2n+2}...\xi_{4n-1,4n})^{1/4}},
\end{eqnarray}
which is same as the $2n$-point correlation function of $O_{1}$ in $t<L-l$ or $L+l<t$. Thus the variation of the $n$-th R\'enyi entropy in the CFT with Neumman boundary condition is
\begin{eqnarray}\label{EarlytimeresultinO2General}
\Delta S_{A}^{(n)}=0.
\end{eqnarray}
In $L-l<t<L+l$ with the limit $\epsilon\rightarrow 0$, due to (\ref{varianceofpoint2}) the leading contribution of (\ref{Key2nBosonO22}) should satisfy following constraints
\begin{eqnarray}\label{constraint1}
i_{2n+1}+i_{2n+2}=0,\ \ i_{2n+3}+i_{2n+4}=0,...,\ \ i_{4n-1}+i_{4n}=0.
\end{eqnarray}
Combining with ${i_{2n+j}}={i_{j}}$  and (\ref{constraint1}), one can obtain
\begin{eqnarray}\label{constraint2}
i_{1}+i_{2}=0,\ \ i_{3}+i_{4}=0,...,\ \ i_{2n-1}+i_{2n}=0.
\end{eqnarray}
In terms of (\ref{varianceofpoint2}), the leading terms of (\ref{Key2nBosonO22}) should also satisfy following constraints
\begin{eqnarray}\label{constraint3}
i_{2}+i_3=0,\ \ i_{4}+i_5=0,...,\ \ i_{2n}+i_{1}=0.
\end{eqnarray}
With these constraints (\ref{constraint1})(\ref{constraint2})(\ref{constraint3}), the correlation function is
\begin{eqnarray}
&&\langle O_2^{\dag}(\xi_1,\bar \xi_1)O_2(\xi_2,\bar \xi_2)...O_2^{\dag}(\xi_{2n-1},\bar \xi_{2n-1})O_2(\xi_{2n},\bar \xi_{2n})\rangle_{UHP}\nonumber \\
&\simeq&  \frac{2}{2^{n}}\frac{1}{(\xi_{23}...\xi_{2n,1})^{1/4}(\xi_{2n+1,2n+2}...\xi_{4n-1,4n})^{1/4}}\label{freen}.
\end{eqnarray}
Putting (\ref{freen}) into (\ref{result1}) with considering Jacobi factor (\ref{DerivitiveGeneralzationfour21})(\ref{DerivitiveGeneralzationfour22}), the variation of the $n$-th R\'enyi entropy is
\begin{eqnarray}\label{LatertimeresultO2General}
\Delta S_{A}^{(n)}=\log 2.
\end{eqnarray}
In CFT with the Dirichlet boundary, the only difference with the Neumann boundary condition is the constraints $i_{2n+j}=i_{j}$ $\to$ $i_{2n+j}=-i_{j}$. One could check that the leading order correlation function is same as the Neumann boundary condition. Thus the $n$-th R\'enyi entropy is not dependent on the choice of boundary condition. We do not repeat here.

\subsection{R\'enyi Entropy in Ising model}
It is natural to ask how about the Ising model which is simplest unitary minimal model. There are three kinds of primary operators, i.e., the identity $\bm{\mathbb{I}}$, the spin operator $\bm{\sigma}$ and the energy operator $\bm{\epsilon}$. There are also two kinds of parity transformation which involve in two kinds of boundary conditions, which correspond to two different parity transformations. One of the parity transformation \cite{CFT} is
\begin{eqnarray}\label{parityIsing1}
\sigma(z,\bar z)\to \sigma(\bar z,z),\ \ \ \ \ \mu(z,\bar z)\to \mu(\bar z,z),
\end{eqnarray}\label{parityIsing2}
where $\bm{\mu}$ is the disorder operator. The other parity transformation \cite{CFT} is
\begin{eqnarray}
\sigma(z,\bar z)\to \mu(\bar z,z),\ \ \ \ \ \mu(z,\bar z)\to \sigma(\bar z,z).
\end{eqnarray}
We would like to study the local excitation by the spin operator $\bm{\sigma}$ with conformal dimension ($h=\frac{1}{16}$, $\bar h=\frac{1}{16}$) in the same setup given in section (\ref{3.1}). The variation of the R\'enyi entropy for the subsystem $-l<x<0$ is given by (\ref{result1}).

The two-point correlation function
\begin{eqnarray}
&&\langle\sigma^{\dag}(\omega_1,\bar \omega_1)\sigma(\omega_2,\bar \omega_2)\rangle_{B_1}=\prod_{i=1}^2(\frac{d\omega_i}{d\xi_i^{'}})^{-h}(\frac{d\bar \omega_i}
{d\bar \xi_i^{'}})^{-\bar h}\langle \sigma^{\dag}(\xi_1^{'},\bar \xi_1^{'})\sigma(\xi_2^{'},\bar \xi_2^{'})\rangle_{UHP}\nonumber \\
&=&\prod_{i=1}^2(\frac{d\omega_i}{d\xi_i^{'}})^{-h}(\frac{d\bar \omega_i}
{d\bar \xi_i^{'}})^{-\bar h}\langle \sigma^{\dag}(\xi^{'}_1) \sigma(\xi^{'}_2) \tilde{\sigma}^{\dag}(\xi^{'}_3) \tilde{\sigma}(\xi^{'}_4) \rangle_{R_2}
\end{eqnarray}
where $\xi_3^{'}\equiv \xi_1^{'*}$,$\xi_4\equiv \xi_2^{'*}$,$\xi_i^{'}=i\omega_i$, $\tilde{\sigma}$ is the field with parity transformation.

The 2-point correlation functions on UHP have already obtained in literature, e.g., \cite{Cardy1}\cite{Dotsenko:1984nm}\cite{Dotsenko:1984ad},
\begin{eqnarray}\label{IsingTwopointfunction}
&&\langle \sigma^{\dag}(\xi^{'}_1) \sigma(\xi^{'}_2) \sigma^{\dag}(\xi^{'}_3) \sigma(\xi^{'}_4) \rangle_{R_2}= (\frac{\xi^{'}_{13}\xi^{'}_{24}}{\xi^{'}_{12}\xi^{'}_{23}\xi^{'}_{14}\xi^{'}_{34}})^{\frac{1}{8}}F_{+}(x^{'}),\nonumber \\
&&\langle \sigma^{\dag}(\xi^{'}_1) \sigma(\xi^{'}_2) \mu^{\dag}(\xi^{'}_3) \mu(\xi^{'}_4) \rangle_{R_2}= (\frac{\xi^{'}_{13}\xi^{'}_{24}}{\xi^{'}_{12}\xi^{'}_{23}\xi^{'}_{14}\xi^{'}_{34}})^{\frac{1}{8}}F_{-}(x^{'}),
\end{eqnarray}
with conformal blocks \cite{BPZ}
\begin{eqnarray}\label{F}
&&F_{+}(x^{'})=\sqrt{\sqrt{1-x^{'}}+1}+\sqrt{\sqrt{1-x^{'}}-1},\text{ }\text{  }\text{and}\nonumber \\
&&F_{-}(x^{'})=\sqrt{\sqrt{1-x^{'}}+1}-\sqrt{\sqrt{1-x^{'}}-1},
\end{eqnarray}
where $x^{'}$ is the conformal cross ratio $x^{'}=\xi^{'}_{12}\xi^{'}_{34}/\xi^{'}_{13}\xi^{'}_{24}$. $F_{+}(x^{'})$ and $F_{-}(x^{'})$ correspond to different boundary conditions respectively.
The leading behavior of the 2-point correlation function in $\epsilon \rightarrow 0$ is
\begin{eqnarray}\label{2pointising}
\langle\sigma^{\dag}(\omega_1,\bar \omega_1)\sigma(\omega_2,\bar \omega_2)\rangle_{B_1}\simeq \frac{\sqrt{2}}{(4\epsilon^2)^{1/8}}.
\end{eqnarray} (\ref{2pointising}) for both boundary conditions.

In $t>L+l$ or $t<L-l$ with $\epsilon \rightarrow 0$, the leading behavior of 4-point correlation function is
\begin{eqnarray}\label{4pointIsing}
\langle \sigma^{\dag}(\xi_1,\bar \xi_1)\sigma(\xi_2,\bar \xi_2) \sigma^{\dag}(\xi_3,\bar \xi_3)\sigma(\xi_4,\bar \xi_4) \rangle_{UHP}&\simeq& \langle \sigma^{\dag}(\xi_1,\bar \xi_1)\sigma(\xi_2,\bar \xi_2)\rangle_{UHP}\langle \sigma^{\dag}(\xi_3,\bar \xi_3)\sigma(\xi_4,\bar \xi_4) \rangle_{UHP}\nonumber\\ &=&  \frac{2}{(\xi_{12}\xi_{56})^{1/8}(\xi_{34}\xi_{78})^{1/8}},\nonumber\\
\end{eqnarray}
where we have used the 2-point function of Ising model on UHP (\ref{IsingTwopointfunction}).
In terms of (\ref{result1}), $\Delta S_{A}^{(2)}$ is
\begin{eqnarray}\label{Isingresultone1}
\Delta S_{A}^{(2)}&\simeq& -\log \Big[(\frac{(\xi_{12}\xi_{34}\xi_{56}\xi_{78})^{2}}{(2\epsilon)^{4}})^{1/16}  \frac{\langle \sigma_2^{\dag}(\xi_1,\bar \xi_1)...\sigma_2(\xi_4,\bar
\xi_4)\rangle_{UHP}}{(\langle \sigma^{\dag}_2(\xi_1,\bar \xi_1) \sigma_2(\xi_2,\bar \xi_2)\rangle_{UHP})^2}\Big]\nonumber \\
&=&-\log \Big[(\frac{(\xi_{12}\xi_{34}\xi_{56}\xi_{78})^{2}}{(2\epsilon)^{4}})^{1/16} \frac{(4\epsilon^2)^{1/8}}{(\xi_{12}\xi_{34}\xi_{56}\xi_{78})^{1/8}}\Big]=0,
\end{eqnarray}

In $L-l<t<L+l$ with the limit $\epsilon \rightarrow 0$, the 4-point correlation function (\ref{4pointIsing}) on UHP can not be factorized directly. We also use the image method, which states that the 4-point correlation function on UHP can be expressed as linear combination of the holomorphic part of conformal blocks of the 8-point correlation function on the full complex plane, with coordinates $\xi_1...\xi_8$. As we know in 2 dimension full complex plane, there are $(n-3)$ independent cross ratios for  $n$-point correlation function. In our case, there are 5 independent cross ratios. Thus the 4-point correlation function on UHP can be expressed by conformal blocks \cite{BPZ}
\begin{eqnarray}\label{UHP-Ising}
&&\langle \sigma^{\dag}(\xi_1,\bar \xi_1)\sigma(\xi_2,\bar \xi_2) \sigma^{\dag}(\xi_3,\bar \xi_3)\sigma(\xi_4,\bar \xi_4) \rangle_{UHP}\nonumber\\&=&(\frac{\xi_{13}\xi_{24}}{\xi_{12}\xi_{23}\xi_{14}\xi_{34}})^{\frac{1}{8}}\sum_b A^bC^b \mathcal{F}[b;x,x_1,x_2,...,x_4],
\end{eqnarray}
where we define the 5 independent conformal ratios $x,x_1,...,x_4$, i.e. $x=\xi_{12} \xi_{34}/\xi_{13}\xi_{24}$, $x_1=\xi_{14} \xi_{\bar{1}\bar{2}}/\xi_{1\bar{1}}\xi_{4\bar{2}},x_2=\xi_{1\bar{1}} \xi_{\bar{2}\bar{3}}/\xi_{1\bar{2}}\xi_{\bar{1}\bar{3}},x_3=\xi_{2\bar{1}} \xi_{\bar{2}\bar{4}}/\xi_{2\bar{2}}\xi_{\bar{1}\bar{4}}, x_4=\xi_{3\bar{2}} \xi_{\bar{3}\bar{4}}/\xi_{3\bar{3}}\xi_{\bar{2}\bar{4}}$, $b$ in conformal blocks $\mathcal{F}$ runs over all the intermediate conformal families and the coefficients $A^p$ are determined by boundary conditions. Here we define $\xi_{\bar{i}}=\bar{\xi}_{i}$. The conformal blocks \cite{BPZ} satisfy the fusion transformation under the braiding operation \cite{Moore:1988uz}\cite{Verlinde:1988sn}\cite{Lewellen:1991tb}, i.e.,
\begin{eqnarray}
\mathcal{F}[b;x,x_1,x_2,...,x_4]=F_{bc} \mathcal{F}[c;1-x,x_1,x_2,x_3,x_4],
\end{eqnarray}
$F_{bc}$ is the fusion matrix. Making the fusion is equal to $\xi_{2}\leftrightarrow \xi_4$. In summary, the leading divergence of the correlation function is related to $c=0$ as identity as follows
\begin{eqnarray}
(\ref{UHP-Ising})=\sum_{b,c}
F_{bc}(\frac{\xi_{13}\xi_{42}}{\xi_{14}\xi_{43}\xi_{12}\xi_{23}})^{\frac{1}{8}}A^b C^b \mathcal{F}[c;1-x,x_1,x_2,...,x_4].
\end{eqnarray}
In terms of (\ref{Varianceofpoitmiddletime}), one can find leading contribution of express (\ref{UHP-Ising}) as following form in $L-l<t<L+l$ with $\epsilon \rightarrow 0$
\begin{eqnarray}
(\ref{UHP-Ising})\simeq F_{00}\langle \sigma^{\dag}(\xi_1,\bar \xi_1)\sigma(\xi_4,\bar \xi_2) \rangle_{UHP} \langle \sigma^{\dag}(\xi_3,\bar \xi_3)\sigma(\xi_2,\bar \xi_4) \rangle_{UHP},
\end{eqnarray} where $0$ stands for the identity operator.
Then the variation of the R\'enyi entropy for $n=2$
\begin{eqnarray}
\Delta S^{(2)}_{A}=-\log F_{00}=\log \sqrt{2},
\end{eqnarray}
where we have used the fact $F_{00}=\frac{1}{\sqrt{2}}$ \cite{Moore:1988ss} in two-dimensional Ising model. Note that the R\'enyi entropy does not depend on the choice of boundary condition in the Ising model, since the leading behavior of $F_{+}$ is the same as one of $F_{-}$  with $\epsilon \rightarrow 0$.

\begin{figure}[h]
\begin{center}
\epsfxsize=8.5 cm \epsfysize=1.5 cm \epsfbox{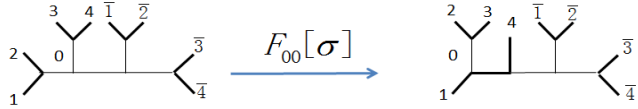}
\end{center}
\caption{Fusion transformation in Ising model. Where $i$ and $\bar{i}$ presented in the figure stands for the position $z_i$ and $\bar{z}_i$ respectively. $F_{00}[\sigma]$ denotes the fusion transformation and $0$ corresponds to the identity operator.}\label{f4-T-4}
\end{figure}

One alternative way to understand the phenomenon is to make use of diagram representation of conformal block as fig.[\ref{f4-T-4}].
Since the behavior of coordinates $\bar \xi_j$ does not change when $L-l<t<L+l$, we only need one time fusion transformation, which is different with \cite{He:2014mwa}.
In the fig.[\ref{f4-T-4}],
\begin{eqnarray}F_{00}[\sigma]=F_{00}\left[
\begin{array}{cc}
 \sigma  & \sigma  \\
 \sigma  & \sigma  \\
\end{array}
\right].\end{eqnarray}
In $L-l<t<L+l$ with $\epsilon \rightarrow 0$, the leading contribution of (\ref{UHP-Ising}) originates from the above conformal block involving in identity operator. Because the fusion factor $F_{00}$ can not be canceled in the ratio (\ref{result1}), $\Delta S^{(2)}_{A}=-\log F_{00}=\log \sqrt{2}$ \cite{Moore:1988ss}.

\subsection{R\'enyi Entropy in General Rational CFTs}
In this subsection, we would like to generalize the analysis to the rational CFTs with a boundary in our previous set-up shown in subsection (\ref{3.1}). In terms of (\ref{result1}), we should know the 2-point correlation on $B_1$ and $2n$-point correlation fuction on UHP as usual.
In generic rational CFTs with a boundary, the 2-point correlation can be expressed as
\begin{eqnarray}\label{2pointfucntiongeneral}
&&\langle O^{\dag}(\omega_1,\bar \omega_1)O(\omega_2,\bar \omega_2)\rangle_{B_1}=\langle O^{\dag}(\xi_1^{'},\bar \xi^{'}_1)O(\xi^{'}_2,\bar \xi^{'}_2)\rangle_{UHP}\nonumber \\
&=&\frac{1}{(\xi^{'}_{13}\xi^{'}_{24})^{2h}}\sum_b A^b C^b \mathcal{F}[b;\xi^{'}],
\end{eqnarray}
where $A^b$ are constants which are determined by the boundary condition, $\xi^{'}_3$, $\xi^{'}_4$, $\xi^{'}$ are given in (\ref{twopointfunction}).
4-point correlation function \cite{Dotsenko:1984nm}\cite{Dotsenko:1984ad} of a primary function $O$ on the $R^2$, which can
be expressed by
\begin{eqnarray}\label{genericfourpoint}
\langle O^{\dag}(z_1,\bar z_1) O(z_2,\bar z_2) O^{\dag} (z_3,\bar z_3)O (z_4,\bar z_4) \rangle_{R_2}=\sum_b\frac{1}{(z_{13}z_{24})^{2h}} C^b \mathcal{F}[b;z]\times c.c.
\end{eqnarray}
In the limit $z\to 0$
\begin{eqnarray}\label{Blocksim}
\mathcal{F}[b;z]=z^{h_b-2h}+...,
\end{eqnarray}
where ``..." stands for higher order terms and $z={z_{12}z_{34}\over z_{13}z_{24}}$ and $h_b$ is conformal dimension of primary class $b$.

In terms of (\ref{Blocksim}), (\ref{2pointfucntiongeneral}) with taking $\xi^{'}\to 0$ is
\begin{eqnarray}
\langle O^{\dag}(\omega_1,\bar \omega_1)O(\omega_2,\bar \omega_2)\rangle_{B_1}&=&\frac{1}{(\xi^{'}_{13}\xi^{'}_{24})^{2h}} A^0 C^0 {\xi^{'}}^{-2h}
\simeq \frac{A^0 C^0}{(4\epsilon^2)^{2h}},
\end{eqnarray}
In $t>L+l$ or $t<L-l$ with $\epsilon \rightarrow 0$, the 2n-point correlation function on UHP is
\begin{eqnarray}\label{2npointfunction}
 &&\langle O^{\dag}(\xi_1,\bar \xi_1)O(\xi_2,\bar \xi_2)...O^{\dag}(\xi_{2n-1},\bar \xi_{2n-1})O(\xi_{2n},\bar \xi_{2n})\rangle_{UHP}\nonumber \\
 &\simeq& \langle O^{\dag}(\xi_1,\bar \xi_1)O(\xi_2,\bar \xi_2)\rangle_{UHP} ...\langle O^{\dag}(\xi_{2n-1}, \bar \xi_{2n-1})O(\xi_{2n},\bar \xi_{2n})\rangle_{UHP}.
\end{eqnarray}
The 2-point correlation on UHP is
\begin{eqnarray} \label{2pointfunctionafterfactor}
\langle O^{\dag}(\xi_{2k-1},\bar \xi_{2k-1})O(\xi_{2k},\bar \xi_{2k})\rangle_{UHP}\simeq \frac{A^0C^0}{(\xi_{2k-1,2k}\xi_{2n+2k-1,2n+2k})^{2h}},
\end{eqnarray}
where $1\le k\le n$, $\xi_{2n+2k-1}\equiv \bar \xi_{2k-1}$.

Taking (\ref{2npointfunction})(\ref{2pointfunctionafterfactor}) (\ref{varianceofpoint1})(\ref{2pointfucntiongeneral})and Jacobi factor (\ref{DerivitiveGeneralzation11})(\ref{DerivitiveGeneralzation12})into (\ref{result1}), one can obtain
\begin{eqnarray}\label{genralresult1}
\Delta S_{A}^{(n)}=0.
\end{eqnarray}

In $L-l<t<l+L$ with $\epsilon \rightarrow 0$, the $2n$-point correlation function on UHP could be written as a linear combination of the holomorphic conformal blocks of the $4n$-point correlation function. We make the following fusions \cite{Moore:1988uz}\cite{Verlinde:1988sn}\cite{Lewellen:1991tb}
\begin{eqnarray}
&&(\xi_1,\xi_2)(\xi_3,\xi_4)...(\xi_{4n-1},\xi_{4n})\to(\xi_2,\xi_3)(\xi_1,\xi_4)...(\xi_{4n-1},\xi_{4n})\nonumber \\
&\to& (\xi_2,\xi_3)(\xi_4,\xi_5)(\xi_1,\xi_{6})...\to...\to (\xi_2,\xi_3)...(\xi_1,\xi_{2n})...(\xi_{4n-1},\xi_{4n}),
\end{eqnarray}
where $\bar \xi_{2n+i}=\xi_i$, $0\le i\le 2n$. With using $n-1$ times fusion transformation as the one shown in fig.[\ref{f4-T-4}], we get
\begin{eqnarray}
&&\langle O^{\dag}(\xi_1,\bar \xi_1)O(\xi_2,\bar \xi_2)...O^{\dag}(\xi_{2n-1},\bar \xi_{2n-1})O(\xi_{2n},\bar \xi_{2n})\rangle_{UHP}\nonumber \\
&\simeq& F_{00}[O]^{n-1} \langle O^{\dag}(\xi_1,\bar \xi_1)O(\xi_{2n},\bar \xi_2)\rangle_{UHP} ...\langle O^{\dag}(\xi_{2n-2}, \bar \xi_{2n-1})O(\xi_{2n-1},\bar \xi_{2n})\rangle_{UHP},
\end{eqnarray}
We could calculate $\Delta S^n_A$ by using (\ref{result1}).
The variation of the $n$-R\'enyi entropy is
\begin{eqnarray}\label{genralresult2}
\Delta S_{A}^{(n)}=-\log F_{00}=\log d_{O},
\end{eqnarray}
where $d_O$ is the quantum dimension \cite{Moore:1988ss} of operator $O$. (\ref{genralresult1}) and (\ref{genralresult2}) are same as the case in rational CFTs living on the full complex plane. We apply the fusion rule of conformal block of $2n$-point function on UHP to obtain the $\Delta S^{(n)}_A$. As we know, the $2n$-point correlation function on UHP can be expressed by linear combination of chiral part of conformal block associated with $4n$-point correlation on full plane. In this subsection, we do not make use of parity transformation containing the boundary information. The boundary data has been encoded in the coefficient of conformal block in this subsection, i.e., $A^p$ in (\ref{genericfourpoint}). We could see that $\Delta S^n_A$ also does not depend on the choice of boundary.

\section{Conclusion and Discussion}
In this paper, we have investigated two kinds of effects on the locally excited states with time evolution. In the first case, we study the locally excited states with thermal effect in low temperature system. We have figured out the thermal correlation which is the same as \cite{Cardy:2014jwa} for the short interval limit. The R\'enyi entropy is equal to summation over the logarithmic of quantum dimension and thermal entropy in low temperature during the time $L-l<t<L+l$. In this paper, we just only confirm that such kind of sum rule is only true for the short interval $l$ in the low temperature limit. We make use of different approach \cite{Herzog:2012bw} to obtain the thermal correction to R\'enyi entropy which can be reduced to \cite{Caputa:2014eta} in low temperature. One can also calculate the R\'enyi entropy in the large interval\cite{Cardy:2014jwa} limit, the higher temperature limit\cite{Chen:2014hta} as well as beyond the leading order of the perturbation (\ref{Expand}). We expected the sum rule relation is still hold in those cases. But one should note that we actually do not consider the back-reaction of the locally excited states to the thermal environment. When the energy of the local excitation is much lower than the thermal environment, it is safe to ignore the back-reaction. But in some special situation we expect the sum rule will break down. It is an interesting topic to consider in the future.

In the second case, we have studied the R\'enyi entropy of local excited states in 2 dimensional CFTs with a boundary. For 2D CFTs with a boundary, to obtain R\'enyi entropy can be converted to obtain the correlation function on UHP by using of conformal transformation technique. As a warm up, the R\'enyi entropy has been calculated with help of image method in the 2D free field theory with a boundary. The R\'enyi entropy is vanishing for operator $O_2$ (\ref{OperatorBoson}) in $t>L+l$ or $t<L-l$ and $\log 2$ in $L-l<t<l+L$, which is the same as previous study \cite{He:2014mwa} in full complex plane without boundary. To confirm this fact, the R\'enyi entropy have been calculated in Ising model and more generic rational CFTs. Although the correlation function and conformal blocks in 2D CFTs with boundary are totally different from the ones in 2D CFT without boundary, we get a same maximal value of the R\'enyi entropy for the rational CFTs without a boundary\cite{He:2014mwa}. In generic 2D rational CFTs with a boundary, we confirm that the maximal value of R\'enyi entropy is the same as the one in 2D rational CFTs without boundary. \cite{Jackson:2014nla} also try to understand the fact which is not contract with that the left- and right-moving chiral sectors are decoupled. \cite{Jackson:2014nla} generalize the result in \cite{He:2014mwa} to irrational CFT, for example,  Liouvile CFT. They found that the left-right entanglement entropy saturates the Cardy entropy.  In terms of standard view, the Cardy entropy counts the microscopic entropy of actual CFT spectrum. The Cardy entropy seems to suggest two chiral sectors are decoupled.
 The authors in \cite{Jackson:2014nla} proposed a pragmatic point of view to reconcile \cite{He:2014mwa} with the fact that there should also be a comparably large EE between the two chiral sectors of CFT. For example, Non-chiral local operators will be left-right entangled. In BCFTs, the two chiral sectors are no longer independent. We have shown some additional examples to confirm the pragmatic perspective.

For general rational CFTs in 2D, the R\'enyi entropy highly relys on the conformal blocks of the theory. The $n$-point correlation functions in 2D CFTs with boundary are related to the holomorphic part of conformal blocks of the $2n$-point correlation functions on the 2D full complex plane. This relationship had been studied by the image method \cite{Cardy1}\cite{Cardy2} very well. More precise relation is that an $n$-point function in the UHP, which is a function of the coordinates $(z_1,,z_n; \bar{z}_1,...,\bar{z}_n)$
behaves under conformal transformations in the same way as the holomorphic factor of a $2n$-point function in the full plane
which depends on $(z_1,...,z_n; z_1^*,...,z_n^*)$, analytically continued to $z_j^*=\bar{z_j}$. In \cite{He:2014mwa}, the time evolution of R\'enyi entropy highly depends on the holomorphic part of conformal block. In 2D CFTs with a boundary, the boundary changes the evolution of the R\'enyi entropy but does not change the value of the R\'enyi entropy, which is closely related to fusion constants in the bulk. Because the behavior of the `image' coordinates (anti-holomorphic coordinates) does not change as the holomorphic coordinates when $L-l<t<L+l$, we get the same R\'enyi entropy as \cite{He:2014mwa}.

\begin{figure}[h]
\begin{center}
\epsfxsize=12.0 cm \epsfysize=4.0 cm \epsfbox{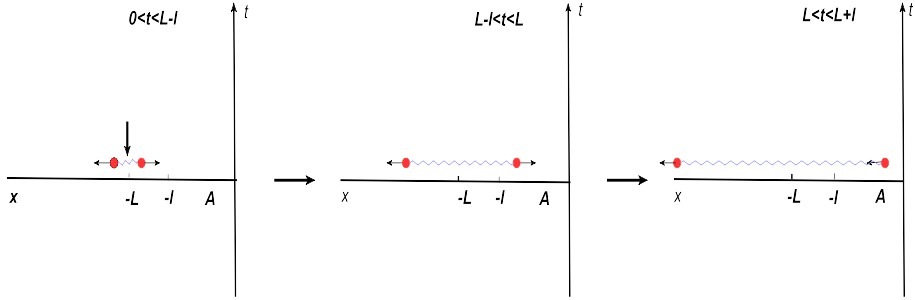}
\end{center}
\caption{To show the time evolution of the entangled quasi-particles pairs. The boundary $x=0$ works as reflector. These red points stand for quasi-particles and the blue wave lines correspond to the entanglement between them.}\label{fig2}
\end{figure}

The boundary introduced here works as the infinite potential barrier for the time evolution of the entangled quasi-particles pairs \cite{Jean-Marie Stphan}\cite{Nozaki:2014hna}\cite{He:2014mwa} triggered by local excitation as shown in fig.[\ref{fig2}].
The the R\'enyi entropy measures the entanglement between the quasi-particles generated by local excitation. After entangled pairs are created at $-L$, the two quasi-particles will propagate in two opposite directions, i.e., left-moving and right moving.  When the right-moving particle enter the interval $-l<x<0$ denoted by $A$, the R\'enyi entropy takes maximal value due to entanglement between two entangled particles. In fig.[\ref{fig2}], the blue wave lines mimic the entanglement of two entangled quasi-particles. When the right-moving quasi-particle reaches the boundary, the quasi-particle will be reflected by the boundary without losing energy. As the calculation in \cite{Caputa:2014vaa} shows the locally excited states carry the energy of  $O(\epsilon^{-1})$. The conformal transformation, $z\to z+\epsilon(z)$ and $\bar z\to \bar z+\bar \epsilon(\bar z)$, should keep the boundary conformal invariant, which lead to the constraint $T=\bar T$ on the boundary. In the Cartesian coordinates, the constraint becomes $T_{xt}=0$, which means that no energy can flow across the boundary. This is main reason the quasi-particle must be reflected no matter what is the conformal invariant boundary condition. In this sense, the boundary change the time evolution of R\'enyi entropy.In our paper we take the scale $\epsilon$ as the minimal scale, and keep the leading order of $\epsilon$ in the calculation. So we must miss some information when $t\sim L$, i.e., the quasi-particle is close to the boundary. In this case, we should make use of bulk boundary correlation functions and boundary structure constants in BCFT to figure out the time evolution of entangled quasi-particles. The next leading order calculation of $\epsilon$ may give us more insight on this point.

\vskip 0.5cm
{\bf Acknowledgement}
\vskip 0.2cm
We are grateful to J. L. Cardy, Mitsutoshi Fujita, Rene Meyer, Masahiro Nozaki, T.~Numasawa, Noburo Shiba, Tadashi Takayanagi and K.~Watanabe for useful conversations and correspondence. We thank Miao Li, Tadashi Takayanagi for their encouragement and support. W. Z. Guo is supported by Postgraduate Scholarship Program of China Scholarship Council. S.H. is supported by JSPS postdoctoral fellowship for foreign researchers and by the National Natural Science Foundation of China (No.11305235).

\end{document}